\documentclass[aps,prb,twocolumn,floatfix]{revtex4}
\usepackage{amsmath,amssymb,epsfig,psfig}

\begin{document}


\title{Thermodynamic and spectral properties of compressed Ce
calculated by the merger\\ of the local density approximation and
dynamical mean field theory}

\author{A. K. McMahan,$^1$ K. Held,$^2$ and R. T. Scalettar$^3$}
\affiliation{$^1$Lawrence Livermore National Laboratory, University
of California, Livermore, CA 94550\\ $^2$Max-Planck-Institut f\"ur
Festk\"orperforschung, D-70569 Stuttgart, Germany\\ $^3$Physics
Department, University of California, Davis, CA 95616}
\date{\today}

\begin{abstract}
We have calculated thermodynamic and spectral properties of Ce
metal over a wide range of volume and temperature, including
the effects of $4f$ electron correlations, by the merger of
the local density approximation and dynamical mean field theory
(DMFT).  The DMFT equations are solved using the quantum Monte
Carlo  technique supplemented by the more approximate Hubbard I
and Hartree Fock methods.  At large volume we find Hubbard split
spectra, the associated local moment, and an entropy consistent
with degeneracy in the moment direction.  On compression through
the volume range of the observed $\gamma$-$\alpha$ transition,
an Abrikosov-Suhl resonance begins to grow rapidly in the $4f$
spectra at the Fermi level, a corresponding peak develops in the
specific heat, and the entropy drops rapidly in the presence
of a persistent, although somewhat reduced local moment.
Our parameter-free spectra agree well with experiment at the
$\alpha$- and $\gamma$-Ce volumes, and a region of negative
curvature in the correlation energy leads to a shallowness in
the low-temperature total energy over this volume range which is
consistent with the $\gamma$-$\alpha$ transition.  As measured by
the double occupancy, we find a noticeable decrease in correlation
on compression across the transition; however, even at the smallest
volumes considered, Ce remains strongly correlated with residual
Hubbard bands to either side of a dominant Fermi-level structure.
These characteristics are discussed in light of current theories
for the volume collapse transition in Ce.\\

\end{abstract}


\maketitle

\section{Introduction}

A number of rare earth metals undergo pressure induced first order
phase transitions with unusually large volume changes of 9--15\%
(for reviews see Refs.~\onlinecite{Benedict93,Holzapfel95,JCAMD}).
Of these transitions the isostructural
$\gamma$-$\alpha$ transition in Ce has received the most
attention.\cite{Ce} It was discovered first, has the largest
volume change (15\% at room temperature), and may also be accessed
entirely at ambient pressure (or in vacuum) by changing the
temperature, thus, for example, allowing thorough spectroscopic
investigation of both phases.  The results of such photoemission
and Bremsstrahlung  studies \cite{Liu92} show a dramatic transfer
of spectral weight to the Fermi energy and the development of a
large peak with its center of gravity slightly above the Fermi
energy when going from the $\gamma$- to the $\alpha$-Ce phase.
Similarly, the optical conductivity is higher in the $\alpha$ phase
where the frequency dependent scattering rate is characteristic
for a Fermi liquid behavior with an effective mass of about
$20\,m_e$.\cite{Eb01} Also the magnetic susceptibility and its
temperature dependence change from a Curie-Weiss like behavior in
the $\gamma$ phase to a Pauli paramagnetic behavior in the $\alpha$
phase. \cite{Ce} Despite these dramatic differences, the number
of $4f$ electrons does not change significantly and is close to
one across the $\gamma$-$\alpha$ phase transition line which ends
in a critical point at $T\!=\!600\pm 50\,$K, \cite{Ce} above which
the $\gamma$- and $\alpha$-Ce phases become indistinguishable.

Notwithstanding the considerable attention, there remains continued
disagreement about the nature of the transition and the $\alpha$
phase.  In general, it is believed that the transition is driven by
changes in the $4f$ electron correlations, though some alternative
theories have been proposed.  Two recent examples of the latter
assume some kind of symmetry breaking in the $\alpha$ phase:
 Eliashberg and Capellmann \cite{Eliashberg98} argue that
$\alpha$-Ce has a symmetry broken distorted structure mainly based
on the observation that the $\alpha$ phase shows large changes
of the compressibility;\cite{Davis64}
Nikolaev and Michel \cite{Nikolaev} propose (hidden) quadrupolar
ordering.  In these theories  a critical endpoint is impossible
because of the symmetry breaking, and there must be, at least,
a second order phase transition line above $T\!\sim\,$600$\,$K,
which disagrees with the common interpretation of the experiment.
\cite{Ce}

The first theory of electronic origin to describe the
$\gamma$-$\alpha$ transition was the promotional model \cite
{PM} which assumed a change in the electronic configuration
from $4f^1(spd)^3$ to $4f^0(spd)^4$. However, it was soon ruled
out by experiment which did not reveal any major change in the
number of $4f$ electrons.  Also band structure calculations
found about one $4f$ electron per Ce atom in the $\alpha$
phase, leading Johansson to propose a Mott transition (MT)
scenario.\cite{JOHANSSON} Similar to the MT of the Hubbard model, the
$4f$ electrons are considered to be localized in the $\gamma$
phase and to be itinerant in the $\alpha$ phase, with this
reduction in the degree of $4f$-electron correlation being
caused by the decrease in the ratio of Coulomb interaction to
kinetic energy under pressure.  In a subsequent analysis based
on these ideas, Johansson {\it et al.}\cite{JOHANSSON2} employed
a standard local density approximation (LDA) calculation for the
$spdf$ electrons in the $\alpha$ phase, while treating the $4f$
electrons as localized $4f^1$ moments decoupled from LDA $spd$
bands in the $\gamma$ phase.  Evidence for the MT scenario to be
correct is taken from the considerable success of LDA calculations
and their generalized gradient improvements for the structural
and volume dependence of the total energy of $\alpha$-Ce-like
phases.\cite{Brooks84,Soderlind98} Additional support appears to
come from orbitally polarized\cite{Eriksson90,Soderlind02,Svane97}
and self-interaction corrected\cite{Svane97,Svane94,Szotek94}
LDA modifications which obtain transitions in Ce and Pr at about
the right pressures. Also LDA+U calculations have  been reported
for one or both Ce phases.\cite{Sandalov95,Shick01}

This MT model appears in conflict with the Kondo volume collapse
(KVC) scenario of Allen and Martin \cite{KVC} and Lavagna {\it et
al.}\cite{KVC2} which is based on the Anderson impurity model.
Both pictures agree that, at the experimental temperatures, the
larger volume $\gamma$ phase is strongly correlated (localized),
has Hubbard split $4f$ spectra, and exhibits a Hund's rule
$4f^1$ moment as reflected in the observed Curie-Weiss magnetic
susceptibility.  But, while the MT scenario then envisages a
rather abrupt transition on compression to a weakly correlated
(itinerant) $\alpha$ phase, in which the Hubbard split bands have
coalesced together near the Fermi level and the $4f^1$ moments
are lost, the KVC picture assumes continued strong correlation
in the $\alpha$ phase with Kondo screening of the $4f$ moments by
the valence electrons.  The signature of this Kondo screening is
a peak in the $4f$ spectra at the Fermi level, the Abrikosov-Suhl
resonance, which lies between the remaining Hubbard-split spectral
weight characteristic of the  local moment.\cite{Liu92} Assuming
a rapid volume dependence of the Kondo temperature, there is Kondo
screening and an Abrikosov-Suhl resonance in the $\alpha$ phase
but not in the $\gamma$ phase since the Kondo temperature of the
$\alpha$ phase is above and that of the $\gamma$ phase below the
typical experimental temperatures.  The strong volume dependence of
the Kondo temperature leads to a region of negative curvature in
the free energy and, thence, a first order transition as dictated
by the common (Maxwell) tangent construction, \cite{KVC,KVC2}
similar to a vapor-liquid transition.  While the KVC model provides
a genuine many-body calculation for Ce, it nevertheless incorporates only
two bands, which has prompted previous attempts at introducing
more orbital realism from LDA.\cite{Noam88,Laegsgaard99}

On the basis of the underlying Hubbard and periodic Anderson
models, it has been argued recently that the MT and KVC scenarios
are rather similar when the important many-body effects are taken
into account.\cite{Held00ab} 
That is, the behavior of the local moment at the MT of the Hubbard model
is not so abrupt, nor is the appearance of a three peak structure in
the density of states unique to the periodic Anderson model.
This fact can be obscured by the
use of static mean-field approximations (including LDA and its
modifications) especially when describing the $\alpha$ phase.

A new approach to describe Ce including both orbital realism and
electronic correlation effects is now available with the recent
merger\cite{LDADMFT1,LDA++,LDADMFT3} of LDA and dynamical mean
field theory (DMFT).\cite{DMFT1,DMFT2} This approach has been
employed by Z\"olfl {\it et al.} \cite{Zoelfl01} who used the
non-crossing approximation (NCA) to solve the DMFT equations
in order to calculate the spectra, Kondo temperatures, and
susceptibilities for $\alpha$- and $\gamma$-Ce.  Independently,
we treated the DMFT equations with the more rigorous Quantum
Monte Carlo \cite{METHOD} (QMC) simulations and reported, as
first results of the present effort, evidence for a Ce volume
collapse in the total LDA+DMFT energy which coincides with
dramatic changes in the $4f$ spectrum.\cite{Held01b} A similar
transition was also described earlier in LDA+DMFT calculations
for Pu.\cite{SAVRASOV} In all three cases, the $f$ spectra showed
Abrikosov-Suhl resonances lying in between residual Hubbard
splitting for the smaller-volume, less-correlated $\alpha$ phases,
in contrast to the LDA results mentioned above which only obtain
the Fermi-level structure.  Related behavior is also observed
for the Mott transition in V$_2$O$_3$, which was studied recently
by LDA+DMFT.\cite{Held01a}

In the present work we extend Ref.~\onlinecite{Held01b}
to lower temperatures, complement it with Hubbard-I
calculations,\cite{H-I,LDA++} and calculate the volume-dependence
of  additional physical quantities including the entropy, specific
heat, total spectrum, orbital occupation, and the magnetic moment.
In  Section~II, the LDA+DMFT technique is briefly described
along with the Hubbard-I approximation and a new and faster
implementation of the QMC treatment which is subsequently validated
against established approaches.  In Section~III, thermodynamic
results, i.e., the energy, specific heat, entropy, and free energy,
are presented over a wide range of volume and temperature and the
signatures for the $\alpha$-$\gamma$ transition are discussed.
We present the volume- and temperature-dependence of the $4f$- and
the valence $spd$-spectrum and compare to experiment in Section~IV.
The $4f$ occupation, local magnetic moment, and related quantities
are given in Section~V.  Finally, the results of this paper are
summarized and discussed in Section~VI.

\section{Theoretical methods}

The results in this paper have been obtained by the LDA+DMFT
method, that is by the merger of the local density approximation
(LDA) and dynamical mean-field theory (DMFT) which was
recently introduced by Anisimov {\it et al.} \cite{LDADMFT1} and
Lichtenstein and Katsnelson \cite{LDA++} (for an introduction see
Ref.~\onlinecite{LDADMFT3}).  The starting point of this method is
a conventional LDA band structure calculation.  Since electronic
correlations are only treated at a mean field level within LDA,
the most important term for electronic correlations, i.e., the
local Coulomb interaction, is added explicitly.  This defines a
multi-band many-body problem which is solved by DMFT. To solve
the DMFT equations, we employ two different implementations
of the quantum Monte Carlo (QMC) technique as well as the
Hubbard-I\cite{H-I,LDA++} (H-I) approximation.  This section
describes the relevant computational details of our calculations.

\subsection{LDA+DMFT approach}
\label{approach}

Scalar-relativistic, linear muffin-tin orbital
LDA calculations\cite{LMTO1,LMTO2} were performed for face centered cubic (fcc) Ce over a
grid of volumes as described elsewhere.\cite{JCAMD} The associated
$(6s, 6p, 5d, 4f)$ one-electron Hamiltonians define $16\times16$
matrices $H^0_{\rm LDA}$, after shifting the $4f$ site energies
to avoid double counting the Coulomb interaction $U_f$ between
$4f$ electrons.  The latter is explicitly taken into account in
the full second-quantized Hamiltonian for the electrons,
\begin{eqnarray}
H = &&\sum_{{\bf k},lm,l' m',\sigma} (H^0_{\rm LDA}({\bf k}))_{lm,l' m'}
\,\hat{c}^\dagger_{{\bf k}\,lm\sigma} \hat{c}^{ }_{{\bf k}\,l' m'\sigma}
\nonumber \\
&&+\; \frac12 \, U_f \!\!\! \sum_{{\bf i},m\sigma,m'\sigma'} \!\!\!\!\!\!\! ^{'} \,
\hat{n}_{{\bf i}fm\sigma}\, \hat{n}_{{\bf i}fm'\sigma'}.
\label{Ham}
\end{eqnarray}
Here, {\bf k} are Brillouin zone vectors, ${\bf i}$ are lattice sites,
$lm$ denote the angular momentum, $\sigma$ is the spin quantum
number, $\hat{n}_{{\bf i}fm\sigma} \equiv \hat{c}^\dagger_{{\bf i}fm\sigma}
\hat{c}^{ }_{{\bf i}fm\sigma}$, and the prime signifies $m\sigma \neq m'
\sigma'$. The many-body Hamiltonian Eq.~(\ref{Ham}) has {\it no}
free parameters since we employed constrained-occupation LDA
calculations to determine $U_f$ and the $4f$ site energy shift for
all volumes considered (see Fig.~5 of Ref.~\onlinecite{JCAMD}
for the values). We did not take into account the spin-orbit
interaction which has a rather small impact on LDA results for Ce,
and also neglected the intra-atomic exchange interaction which
has only an effect if there are more than one $4f$-electrons on
a Ce atom.

The DMFT maps the lattice problem Eq.~(\ref{Ham}) onto the
self-consistent solution of the Dyson equation
\begin{equation}
G_{\bf k}(i\omega)=\left[ \;i\omega I+\mu I
-H_{{\rm LDA}}^{0}({\bf k})
-\Sigma(i\omega)I_f\right]^{-1} ,
\label{Green}
\end{equation}
and a seven-orbital (auxiliary) impurity problem defined by the bath Green function
\begin{equation}
{\cal G}(i\omega)^{-1}=
 \left(\frac{1}{7N} \sum_{\bf k}
Tr\{G_{\bf k}(i\omega)I_f\}\right)^{-1}\!+\Sigma(i\omega).
\label{impurity}
\end{equation}
Here $I$ is the unit matrix, $I_f \equiv
[\delta_{lf}\delta_{l^\prime f} \delta_{mm^\prime}]$ projects
onto the seven $f$-orbitals, $\mu$ is the chemical potential,
${\rm Tr}$ denotes the trace over the orbital matrix, and $N$ is
the number of {\bf k} points ($N\!=\!2048$ for $T\leq 0.4\,$eV and
$N\!=\!256$ for $T>0.4\,$eV).  Within the LDA, there is a minor
crystal-field splitting of the seven $4f$ orbitals. However, in
Eq.~(\ref{impurity}) we average over the seven $4f$ orbitals, i.e.,
we treat them as degenerate in the auxiliary impurity problem.
Consequently the DMFT self-energy is diagonal $\Sigma(i\omega)I_f$,
at least in the paramagnetic phase studied.  The impurity problem
is solved with one of the methods described in the following two
sections and  generates a self-energy $\Sigma(i\omega)$. This
self-energy gives a new Green function in Eq.~(\ref{Green})
and thus a new impurity problem and so on, iterating to
self-consistency (for more details see Refs.~\onlinecite{DMFT2}
and \onlinecite{LDADMFT3}).  In this self-consistency cycle,
the chemical potential $\mu$ of  Eq.~(\ref{Green}) is adjusted so
that the total number electrons described by Eq.~(\ref{Ham}) is
$n_f\!+\!n_v=4$ per Ce site.  Here, the number of $4f$ electrons
$n_f$, and similarly the number $n_v$ of valence (i.e., $spd$)
electrons, may be obtained from the lattice Green function
\begin{equation}
n_f \! =\! \frac{T}{N} \sum_{n {\bf k} \sigma}
{\rm Tr} \left[G_{\bf k}(i\omega_n)I_f\right] e^{i\omega_n 0^+} \, ,
\label{nf}
\end{equation}
where $T$ is the temperature and $\omega_n=(2n\!+\!1)\pi T$
are the Matsubara frequencies.  
To obtain the physically relevant 
Green function $G(\omega)$, we employ
the maximum entropy method \cite{MEM} for the 
analytic continuation to real frequencies $\omega$.

In principle the LDA and DMFT parts of the calculation should be
mutually self-consistent, with DMFT changes in orbital occupations
(especially $n_f$) feeding back into a new $H^0_{\rm LDA}({\bf k})$
and $U_f$, as argued by Savrasov and coworkers.\cite{SAVRASOV}
Certainly the constrained occupation calculations used to fix
$U_f$ and the $4f$ site energy in $H^0_{\rm LDA}({\bf k})$ should
not be impacted, as they are intended to be valid over the range
$0\!<\!n_f\!<\!2$.  These calculations provide what are in effect
the screened Coulomb energies for 0, 1, and 2 $f$ electrons per
site, which covers this range according to what fraction of the
sites are at one or another of the various occupations.\cite{JCAMD}
However, differences between the DMFT $n_f$ and the LDA $n_f$
could, if the former were fed back into the LDA, change the
position of $4f$ level slightly, and with that the extension of
the $4f$ wavefunction, and thus the $f$-valence hybridization.
It is simply not known at this point if such effects are
important, although we note that DMFT(QMC) and LDA-like (see
Sec.~\ref{energy}) solutions of Eq.~(\ref{Ham}) generally yield
values of $n_f$ within 10\% of one another.  The additional cost on
top of the already very expensive LDA+DMFT(QMC) method also makes
such additional self-consistency impractical in the present case.

\subsection{Hubbard-I approximation}
\label{H-Isec}

In the large-volume limit where intersite hybridization vanishes,
the auxiliary impurity problem is simply the isolated atom,
i.e.,  ${\cal G}(i\omega)= 1/(i\omega+\mu-\varepsilon_f)$ where
$\varepsilon_f$ is the $4f$ site energy.  In this limit the exact
self-energy is known and may, at finite volumes, be used as the
Hubbard-I (H-I) approximation\cite{H-I,LDA++}:
\begin{eqnarray}
\Sigma^{\rm at}(i\omega) &\!=\!& i\omega + \mu_{\rm at} - 
[G^{\rm at}(i\omega)]^{-1} ,
\label{HI-1}\\
G^{\rm at}(i\omega) &\!=\!& \sum_{j\!=\!1}^{14}
\frac{w_j(\mu_{\rm at},T)}{i\omega + \mu_{\rm at} - (j\!-\!1)U_f} ,
\label{HI-2}\\
n_f^{\rm at} &\!=\!& 14T \sum_n G^{\rm at}(i\omega_n)
e^{i\omega_n 0^+} ,
\label{HI-3}
\end{eqnarray}
where $\varepsilon_f$ has been absorbed into $\mu_{\rm at}$, which
is set at each iteration in such a way that $n_f^{\rm at}$ of Eq.~(\ref{HI-3})
equals the current $n_f$ of Eq.~(\ref{nf}).  The positive weights
$w_j$ for transitions between $j\!-\!1$ and $j$ electrons are given by
\begin{eqnarray}
w_j &\!=\!& [j v_j + (15\!-\!j)v_{j\!-\!1}]/
(14\sum_{l=0}^{14} v_l) \,,
\label{HI-4}
\end{eqnarray}
where $v_j$ are Boltzmann weights for having $j$ electrons on the atom
\begin{eqnarray}
v_j &\!=\!& \frac{14!}{j!(14-j)!}\exp\left[-\{\mbox{$\frac12$}
j(j-1)U_f - j\mu_{\rm at}\}/T\right] .
\label{HI-5} 
\end{eqnarray}

Our DMFT(H-I) procedure is in fact also correct at {\it all}
volumes in the high-temperature limit.  Noting that the $w_j$'s
sum to unity, one can see that
\begin{equation}
\Sigma^{\rm at}(\infty) = \frac{13}{14} U_f \sum_{j\!=\!0}^{14} j v_j /
\sum_{j\!=\!0}^{14} v_j = \frac{13}{14} n_f U_f \, ,
\end{equation}
since we always choose $\mu_{\rm at}$ so that $n_f^{\rm
at}\!=\!n_f$.  This is the paramagnetic Hartree-Fock value,
which is also the correct high-temperature limit since the
$\omega_n\propto T$, and only the high-frequency tail of the
self-energy is of importance.

\subsection{QMC simulations}

Our main approach to solve the DMFT impurity problem is the
numerical QMC technique. We use two implementations which differ
mainly by the Fourier transformation between the Matsubara
frequency representation employed in the Dyson equation
Eq.~(\ref{Green}) and the imaginary time representation
employed for the QMC simulation of the impurity problem
Eq.~(\ref{impurity}).  Within QMC, the imaginary time interval
$[0,\beta]$ ($\beta=1/T$) is discretized into $\Lambda$ Trotter
slices of size $\Delta \tau=\beta/\Lambda$.  Since there are $91$
Ising fields {\it per} time slice, the number of time slices which
are computationally manageable in the QMC is seriously restricted.
Thus, if one employs  a discrete Fourier transformation between
${\cal G}(i \omega_n)$ at a finite number of $\Lambda$ Matsubara
frequencies and ${\cal G}(\tau_l), \tau_l\!=\!1\Delta\tau,\dots,\Lambda
\Delta \tau$, the resulting Green function oscillates considerably
around the correct ${\cal G}(\tau)$. To overcome this shortcoming,
Ulmke and coworkers \cite{Ulmke} suggested using a smoothing
procedure which replaces ${\cal G}(i \omega_n) \rightarrow
\tilde{\cal G}(i \omega_n)$, after calculating the auxiliary ${\cal
G}(i \omega_n)$ via Eq.~(\ref{impurity}), where
\begin{equation}
\tilde{\cal G}(i \omega_n) \equiv \frac{\Delta\tau}
{1-\exp[-\Delta \tau/{\cal G}(i \omega_n)]} \, .
\label{smoothing}
\end{equation}
It is $\tilde{\cal G}(i \omega_n)$ that is Fourier transformed to
imaginary time ${\cal G}(\tau_l)$, and once the QMC simulations
of the Anderson impurity problem have yielded the output
$G(\tau_l)$, the process is reversed:
The   Fourier transform of $G(\tau_l)$,
$\tilde{G}(i \omega_n)$, yields $G(i\omega_n)$ from the
inverse of Eq.~{\ref{smoothing}.  The new self-energy is then
$\Sigma(i\omega_n)= {\cal G}(i\omega_n)^{-1}-G(i\omega_n)^{-1}$.

This approach generates smooth Green functions $G(\tau_l)$
and reproduces the correct $\Delta\tau\rightarrow 0$ limit.
We use it in one implementation of the QMC algorithm, referred
to as QMC$_1$ in the following.  Other approaches employed in
the literature are to fit splines to $G(\tau_l)$ and, thus,
to use more support points than $\Lambda$ to do the Fourier
transformation  \cite{DMFT2} or to extend the Matsubara frequency
sums by employing the iterated perturbation theory result at
high frequencies.\cite{METHOD}  Most results of our paper
were obtained by yet another QMC implementation (QMC$_2$) which
uses a different way to Fourier transform and which seems to
be less sensitive to statistical noise.  As this modification
is new, it is described in some detail in Section \ref{QMC2}
and validated in Section \ref{validation}.

\subsubsection{Modified QMC implementation}
\label{QMC2}

In the implementation QMC$_2$, we use a constrained fit
\begin{equation}
G(\tau) = \sum_i w_i f_i(\tau) .
\label{fit}
\end{equation}
to the output QMC impurity Green function $G(\tau_l)$, in
order to accomplish the Fourier transform to $G(i\omega_n)$
for $n=-\frac12 N_\omega,\dots,\frac12 N_\omega\!-\!1$ with
$N_\omega > \Lambda$.  The basis functions are $f_i(\tau)=
-e^{-\varepsilon_i\tau}/(e^{-\beta \varepsilon_i}\!+\!1)$ and  have
Fourier transforms $f_i(i\omega)=1/(i\omega\!-\!\varepsilon_i)$.
At real frequencies, Eq.~(\ref{fit}) corresponds to a set of
$\delta$-functions with different spectral weights $w_i$, and
is capable in the limit of an infinite set of basis functions
of reproducing any given spectrum.  In contrast to a  spline-fit
where every fit-coefficient is determined by the local behavior in
an imaginary time interval, in our approach every fit-coefficient
is determined by the local behavior in frequency space.

The constraints to the fit Eq.~(\ref{fit}) are $w_i\ge 0$, $G(0^+)$
is precisely the QMC value, $G(0^+)\!+\!G(\beta^-)\!=\!-1$, and
$\frac{d}{d\tau}G(0^+) \!+\!\frac{d}{d\tau}G(\beta^-)\!=\!{\rm
g}_2$, where $\beta^-\equiv \beta\!-\!0^+$ and ${\rm g}_m$ is
the $(i\omega)^{-m}$ high-frequency moment of $G(i\omega)$.
For the last constraint, ${\rm g}_2$ is obtained from the
relation $G^{-1}(i\omega)={\cal G}^{-1}(i\omega)-\Sigma(i\omega)$
which implies ${\rm g}_2=g_2+{\rm s}_0$, where these are the
indicated moments of $G(i\omega)$, ${\cal G}(i\omega)$,
and $\Sigma(i\omega)$, respectively.  Here, $g_2$ is
known as ${\cal G}$ is input to the QMC, and we take
$s_0 = \Sigma(i\omega\!=\!\infty) = (13/14)n_f U_f$ with
$n_f=14[1\!+\!G(\tau\!=\!0^+)]$ for the present paramagnetic
case.\cite{s0}

Typically we use grids of $\Lambda/4$ equally spaced
$\varepsilon_i$, and optimize the agreement with the QMC data as
a function of the centroid and width of these grids, in each case
systematically eliminating basis functions for a given grid which
would otherwise yield negative $w_i$.  Because the QMC expense
increases as $\Lambda^3$, we are forced to execute fewer Monte
Carlo sweeps for the largest $\Lambda$'s, and the statistics become
less good than for smaller $\Lambda$'s.  However, the constraint
$w_i\!\ge\! 0$ still seems to provide a sensible interpolation
through the statistical noise, although this has the consequence
that the number of surviving positive $w_i$ increases more slowly
than $\Lambda$. Nonetheless, we see a systematic evolution as a
function of $\Lambda$ and extrapolations $\Lambda\!\rightarrow
\!\infty$ agree with large-volume and high-temperature limits
(Hubbard-I) and the QMC$_1$ (see Section \ref{validation}).  Note
that while we find the fit Eq.~(\ref{fit}) to be very useful for
functional behavior along the imaginary time and frequency axes,
and for integral quantities such as $n_f$ and the total energy,
it is not useful in practice for directly obtaining real frequency
behavior in the presence of typical QMC statistical uncertainties.
The maximum entropy method is far superior here as it folds these
uncertainties into calculation of the spectra.\cite{MEM} 

In order to accelerate the convergence of our DMFT(QMC$_2$)
method we carry out cheap iterations on the constant part of the
self-energy in between each expensive QMC iteration.  That is,
we subtract a constant Hartree-Fock contribution\cite{s0}
from the  QMC self energy: $\Delta\Sigma(i\omega)=\Sigma^{\rm
QMC}(i\omega)- (13/14)n^{\rm QMC}_f U_f$ where $n^{\rm
QMC}_f=14[1\!+\!G(\tau\!=\!0^+)]$.  Following every QMC
cycle, then, one has $\Sigma(i\omega)=(13/14)n_f U_f +
\Delta\Sigma(i\omega)$ in Eq.~(\ref{Green}) which is iterated
to self-consistency with $n_f$ from Eq.~(\ref{nf}), while
keeping $\Delta\Sigma$ fixed.  The resultant values of $n_f$
and $n^{\rm QMC}_f$ agree within statistical uncertainties.
These uncertainties can be significantly smaller for $n_f$ than
for $n^{\rm QMC}_f$ at the smallest volumes.

We find $G(\tau)$ to converge quickly as a function of QMC$_2$
iteration for all $\tau$ at small volume, and for $\tau$ close to
$0$ and $\beta$ at large volume.  For intermediate $\tau$ at large
volume and low temperature, however, where $G(\tau)$ is generally
quite small, convergence appears to result from the average of
frequent small values of $G(\tau)$ with occasional large values as
the Ising configurations are sampled, with the large-volume atomic
limit approached by the latter becoming statistically unimportant.

In order to improve the statistics given this behavior,
we have chosen to include sweeps from all previous QMC
iterations (excluding warm up sweeps) along with the
new sweeps in $G_i^{\rm new}(\tau_l)$ in arriving at the
QMC$_2$ result for iteration $i$: $G_i(\tau_l)=[G_i^{\rm
new}(\tau_l)+(i\!-\!1)G_{i\!-\!1}(\tau_l)]/i$.  Note that
the warm-up sweeps themselves are already started with a
reasonable self-energy, such as a converged DMFT(H-I) result or
a DMFT(QMC$_2$) result for another $\Lambda$.  We have tested
this treatment at both small and large volumes by starting anew
at $i\!=\!1$ from the converged DMFT(QMC$_2$) self-energy,
and have found agreement with the previous results to within
statistical uncertainties.  

We used $10,000$ sweeps per QMC iteration for $\Lambda\!=\!80$,
decreasing systematically to $1,000$ for $\Lambda\!=\!256$, and
carried out from $20$ to over 100 QMC iterations for each $T$,
$V$ point.  At small $V$ even at $T\!=\!0.054\,$eV we found the
DMFT(QMC$_2$) energy to settle down generally after a few QMC$_2$
iterations to maximal excursions of about $\pm 0.02\,$eV ($\pm
0.05\,$eV) for $\Lambda\!=\!80$ (256), with the root-mean-square
uncertainties much smaller.  Such benign behavior extends to
increasingly large volumes at higher $T$, where these DMFT(QMC$_2$)
results begin to agree closely with DMFT(H-I).  At low temperature,
the scatter in our measurements as a function of iteration grows
as volume is increased, especially in the transition region and
beyond; however, the Trotter corrections also diminish here so
there is less need for larger $\Lambda$.  

Finally, we turn to the issue of performing the Fourier transform
from imaginary time to Matsubara frequencies.  The virtue of the
fit Eq.~(\ref{fit}) is that it decouples $\Lambda$ and $N_\omega$
allowing manageable QMC costs (smaller $\Lambda$) and yet accurate
kinetic energies (larger $N_\omega$).  Most of our DMFT(QMC$_2$)
calculations took $N_\omega\!=\!256$ for $T\!\ge\!0.054$ eV
and $N_\omega\!=\!512$ for $T\!=\!0.027$ eV.   In the course of
this work we realized that there is a volume dependence to the
error in the kinetic energy from the Matsubara cutoff, and while the
$N_\omega\!=\!256$ choice at $T\!=\!0.054$ eV leads to a small
$0.04$ eV error in the vicinity of the transition, it becomes
more significant, $0.11$ eV, at the smallest volumes considered.
Since our DMFT(H-I) and DMFT(QMC$_2$) codes have identical kinetic
energy treatment, we used the former to correct the present
DMFT(QMC$_2$) results to effective values of $N_\omega$ four times
those just noted, which should give better than $0.01$ eV accuracy
at all volumes.  We verified this by selected DMFT(QMC$_2$)
tests with the larger $N_\omega$.  Note that this kinetic energy
treatment includes (and the cited errors reflect) an approximate
evaluation of the full infinite Matsubara sum.  Specifically,
we approximate the high-frequency behavior of a quantity
$F(i\omega)$ by $F_0(i\omega)\!=\!w_1/(i\omega\!-\!\varepsilon_1)+
w_2/(i\omega\!-\!\varepsilon_2)$, with parameters chosen to
reproduce its $1/(i\omega)^m$ moments for $m\!=\!1$--$4$.
Then we approximate the infinite Matsubara sum on $F(i\omega)$
by the analytic result for the infinite sum over $F_0$ plus the
finite sum over the difference $F\!-\!F_0$.

\subsubsection{Validation}
\label{validation}

Here we validate the new faster QMC$_2$ algorithm
of Sec.~\ref{QMC2}, used for much of the low-temperature
thermodynamic results in this paper, against QMC$_1$ which employs
the Ulmke-smoothing.  Such validation involves extrapolation to the
limits $N_\omega$,\,$\Lambda\rightarrow\infty$, where the  QMC$_1$
approach should provide exact results.  Errors which vanish in
these limits include those arising from truncation of Matsubara
sums (finite $N_\omega$), and from the Trotter approximation
(finite $\Lambda$).

Figure \ref{kineng} compares the kinetic DMFT  energy (see
Section~\ref{energy} for details of  its calculation)
obtained by QMC$_2$ and  QMC$_1$ as a function of
$\Delta\tau\!=\!\beta/\Lambda$, at a temperature $T\!=\!0.54$ eV
and atomic volume $V\!=\!16.8$ \AA$^3$.   (Note that the $\Delta
\tau$ dependence is largest at small volumes, as we shall discuss
further.) The line with
open circles shows the QMC$_1$ results with Matsubara sums
truncated after $N_\omega\!=\!\Lambda$ frequencies under the
application of Ulmke's smoothing procedure Eq.~(\ref{smoothing}).
Those with squares and open triangles show the results when these
sums are extended to $N_\omega=\infty$ using the Hartree-Fock
(HF) Green function at high Matsubara frequencies; that is,
using Eq.~(\ref{Green}) with $\Sigma\!\rightarrow\!\Sigma_{\rm
HF}\!=\!(13/14)n_fU_f$ for $\pm\omega = \Lambda\pi
T,\dots,\infty$.\cite{HFtails} In the first case (squares) the
current chemical potential $\mu$ and $n_f$ were used to define
$\Sigma_{\rm HF}$.  In the second case (open triangles) the whole
procedure was made self-consistent: From $n_f$, we calculate
$\Sigma(i\omega)\!=\!(13/14)n_fU_f\!+\!\Delta\Sigma(i\omega)$
for all frequencies at a fixed $\Delta\Sigma(i\omega)$ which is
defined in the previous section.  This $\Sigma(i\omega)$ yields a
new $n_f$ via Eq.~(\ref{nf}), and so on until convergence.  As can
be seen, the dependence on $\Delta\tau$ is greatly reduced, as is
also the case for the QMC$_2$ implementation of Section~\ref{QMC2}
(filled circles) which also uses this self-consistent treatment
of the HF part of the the self-energy. To avoid a large $\Delta
\tau$ error, the large frequency part of the self-energy and
especially the constant HF part is important to the energy,
and must be self-consistently correct.
\begin{figure}[tb]
 \includegraphics[width=3.2in]{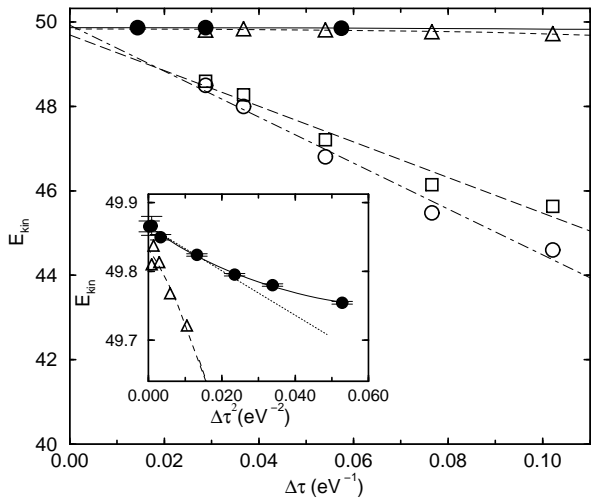}
\caption{ Extrapolation $\Delta \tau\!\rightarrow\! 0$ of the
kinetic energy at $T\!=\!0.54\,$eV and $V\!=\!16.8\,$\AA$^3$,
using the QMC$_2$ (filled circles) and the QMC$_1$ algorithm
(open circles, squares, and triangles; differences are due to
whether and how Hartree-Fock results for the high-frequency
tails of self-energy are included, see text). The lines
show the extrapolations through the QMC data yielding $E_{\rm
kin}(\Delta \tau\!=\!0) = 49.888 \pm 0.003\,$ eV (filled circles)
and $49.853\,\pm 0.022\,$eV (open triangles), both with a mixed
quadratic and cubic fit; and $49.944 \pm  0.271\,$ eV (open circles)
and $49.713\,\pm0.305\,$eV (open squares), both with a linear fit.
The results agree within twice the above standard deviation and,
thus, validate, the QMC$_2$ algorithm.  The inset shows the two
upper curves (filled circles and open triangles) as a function
of $\Delta \tau^2$ over an expanded $\Delta \tau$ range. For the
QMC$_2$ (filled circles) it also compares the mixed quadratic/cubic
fit (solid line) with a purely quadratic fit to the  data points
which fulfill $\Delta \tau U_f/2\leq 0.4$ (dotted line).   
\label{kineng}}
\end{figure}

The inset of Figure \ref{kineng} shows the top two
curves in an expanded scale versus $\Delta\tau^2$, which
indicates $0.035$ eV agreement between the two QMC methods
in the limit $\Delta\tau\!\rightarrow\!0$.  We believe that
effects of Matsubara cutoff are largely eliminated here and
that Trotter errors predominate in these curves, which are
expected to be of leading order $\Delta\tau^2$, at least
for the QMC calculation of lattice models without the DMFT
self-consistency complication.\cite{troterr,DMFT2} In order to
keep the Trotter errors under control, it is recommended that
$\Delta\tau$ be constrained to $\Delta\tau U_f/2<1$,\cite{DMFT2}
and we have done so in this work.  In fact at small volumes
where the Trotter corrections are the largest, we would find
the need to use three terms over this full range, with both
$a+b\Delta\tau^2+c\Delta\tau^3$ and $a+b\Delta\tau+c\Delta\tau^2$
providing reasonable fits to the energy.  We find the ratio
$c/b$ to be significantly smaller for the first choice, which
is consistent with the expectations of a leading $\Delta\tau^2$
dependence.  In our DMFT(QMC$_2$) calculations we have therefore
chosen to use the two term fit $a+b\Delta\tau^2$, however,
over a reduced range.  The dotted line in the inset of Fig.\
\ref{kineng}, for example, suggests that $\Delta\tau U_f/2\leq 0.4$
might be a reasonable range for this fit, given $U_f\!=\!5.05\,$eV
for the volume in the figure.  The two-term fit also makes more
sense in the volume range of the transition, where the Trotter
corrections are smaller, but there is also more scatter as a
function of $\Delta\tau$.

In DMFT(QMC$_2$) calculations for the whole volume grid we have
used at least 1, 2, and 3 $\Delta\tau$ values for temperatures
greater than, equal to, and less then $0.544\,$eV, respectively.
In the first case it is easy to take $\Delta\tau$ so small that
really no extrapolation is needed, or maybe one other value as a
spot check at the smallest volume.  At $T\!=\!0.054\,$eV (632 K)
we used $\Delta\tau U_f/2=0.417$, 0.334, and 0.209.\cite{dtauU} Our
calculations at $T\!=\!0.027\,$eV (316 K) were limited by expense
to systematically larger values, $\Delta\tau U_f/2=0.667$, 0.477,
and 0.334, so that extrapolations to $\Delta\tau\!=\!0$ are more
uncertain.  Even the smallest $\Delta\tau$ here, which corresponds
to $\Lambda\!=\!320$, leads to a $\Delta\tau^2$ value that is
2.6 times larger than its counterpart at $T\!=\!0.054\,$eV.
Fortunately, we see every indication that our electronic
Hamiltonian is already very close to its low-temperature limit
by $T\!=\!0.054\,$eV (632 K), as these total energies agree with
those at $T\!=\!0.027\,$eV within their error bars at the same
{\it finite} $\Delta\tau$ values.  The $\Delta\tau\!\rightarrow\!0$
extrapolations are more benign for $n_f$ and $d$ which also agree
well for the two temperatures.  Accordingly, we do not display
the $T\!=\!0.027\,$eV results in this paper, but do comment on
the agreement between the two temperatures as specific quantities
are presented.

We have alluded earlier to the fact that the Trotter approximation
errors get larger at smaller volume in the present work.
This makes sense as these are related to the commutator of the
kinetic and potential energies, and should thus depend on the size
of the hybridization, which gets larger as volume is reduced.
We find no discernible dependence of the energy on $\Delta\tau$
for volumes in the $\gamma$ phase of Ce for the range of $\Lambda$
investigated, but that for the smaller volume $\alpha$ phase,
we find $dE/d\Delta\tau^2$ to become significant and to increase
in magnitude with decreasing $V$.  Since the $\gamma$--$\alpha$
transition is intrinsically related to the growing importance
of hybridization versus the Coulomb interaction as volume is
decreased, this behavior is perhaps not surprising, although
the effect turns on rather abruptly, appearing as almost another
signature of the transition.
Similar behavior has been seen for the periodic
Anderson model.\cite{HUSCROFT,PAIVA}

\subsection{Calculation of the LDA+DMFT energy}
\label{energy}
There are several possible expressions for the DMFT total energy
per site, depending on whether the potential energy is obtained
using the self energy
\begin{eqnarray}
E_{\rm DMFT}&\! =\!& \frac{T}{N} \sum_{n {\bf k} \sigma}
{\rm Tr}\left[\{ H^0_{\rm LDA}({\bf k})
+ \mbox{$\frac12$}\Sigma(i\omega_n)\} \right.
\nonumber \\
&& \times \left. G_{\bf k}(i\omega_n)\right] e^{i\omega_n 0^+} \, ,
\label{Eng1}
\end{eqnarray}
or from a thermal average of the interaction in Eq.\ (\ref{Ham})
\begin{equation}
E_{\rm DMFT}\! =\! \frac{T}{N} \sum_{n {\bf k} \sigma}
{\rm Tr}\left[{ H}^0_{\rm LDA}({\bf k}) { G}_{\bf k}(i\omega_n)\right]
e^{i\omega_n 0^+} + U_f \, d \, .
\label{Eng2}
\end{equation}
In the latter expression, 
\begin{equation}
d \!=\! \frac{1}{2N}
\sum_{\bf i} \sum_{m\sigma,m'\sigma'}^{\;\;\;\;\;\;\;\;\prime} \langle
\hat{n}_{{\bf i}fm\sigma}\, \hat{n}_{{\bf i}fm'\sigma'}\rangle \, 
\label{double}
\end{equation}
is a generalization of the one-band double occupation
for multi-band models, which may be calculated directly
in the QMC presuming the site average is given by the
associated impurity problem.  If  we were using the {\it
exact}, {\bf k}-dependent self-energy in these equations,
Eq.~(\ref{Eng1}) would be equivalent to  Eq.~(\ref{Eng2}) and to
the Galitskii-Migdal\cite{GM} expression for the total energy.
We find Eq.~(\ref{Eng2}) to be far superior in the present
LDA+DMFT(QMC) calculations in regard to low-temperature stability
and agreement with known limits, possibly not surprising in that it
takes a thermal expectation of the {\it true} Coulomb interaction
for the problem.  We use Eq.~(\ref{Eng1}) for the LDA+DMFT(H-I)
energy in  preference
to Eq.~(\ref{Eng2}) with a purely atomic calculation of $d$.
However, it should be noted that for vanishing intersite
hybridization at large volume, as well as at high temperatures,
the H-I self-energy is exact, and indeed we find agreement between
QMC and H-I results for $E_{\rm DMFT}$ in these limits.

To evaluate the total LDA+DMFT energy $E_{\rm tot}(T)$ including
all core and outer electrons, we add a correction to the
paramagnetic all-electron LDA energy $E_{\rm LDA}(T)$
\begin{equation}
E_{\rm tot}(T)=E_{\rm LDA}(T)+E_{\rm DMFT}(T)-E_{\rm mLDA}(T) \, ,
\label{Etoteqn}
\end{equation}
which consists of the DMFT energy $E_{\rm DMFT}(T)$ less an
LDA-like solution of the same many-body model Hamiltonian
Eq.~(\ref{Ham}), thus ``model LDA'' or $E_{\rm mLDA}(T)$.
The latter is achieved by a self-consistent solution of
Eqs.~(\ref{Green}) and (\ref{nf}) for $n_f$ taking a self-energy
$\Sigma_{\rm mLDA} = U_f(n_f-\frac12)$. From this, the kinetic
energy is calculated by the first term of Eq.~(\ref{Eng2}) and
the potential energy by $\frac12 U_f n_f(n_f-1)$.  Note that
while all of these expressions are explicitly temperature
dependent,\cite{ELDAT} the present calculations are electronic
only and do not attempt to add lattice-vibrational contributions.
Estimates of these contributions are similar for the $\alpha$-
and $\gamma$-Ce phases, however, and appear to have little impact
on the phase diagram.\cite{JOHANSSON2}

Finally, one virtue of the Hamiltonians Eq.~(\ref{Ham})
 is that it is possible to reach high-temperature limits
where the entropy is precisely known.  One may then calculate
the entropy from the DMFT energy
\begin{equation}
S_{\rm DMFT}(T) \!=\! S_\infty - k_{\rm B} \int_{T}^{\infty} dT^\prime
\frac{1}{T^\prime} \frac{dE_{\rm DMFT}(T^\prime)}{dT^\prime} \, ,
\label{Seqn}
\end{equation}
where $S_\infty \!=\! k_{\rm B}[M\ln{M} - n\ln{n} -
(M\!-\!n)\ln{(M\!-\!n)}]=12.057\,k_{\rm B}$, for $M\!=\!32$ states
and $n\!=\!4$ electrons per site for Ce.

\section{Thermodynamics}
\label{TD}

In this section we consider thermodynamic properties of Ce,
more specifically the energy, specific heat, entropy, and free
energy, over a wide range of volume $V$ and temperature $T$.
Intercomparison of the Hartree-Fock (HF), DMFT(H-I), and DMFT(QMC)
methods to solve the effective LDA Hamiltonian Eq.~(\ref{Ham})
here serves to validate all three calculations in limits where
they should and do give the same answers, and also to point
out shortcomings of the more approximate techniques elsewhere.
Then we turn specifically to the $\alpha$--$\gamma$ transition in
Ce, and use the total energy, Eq.~(\ref{Etoteqn}), and entropy,
Eq.~(\ref{Seqn}), to present evidence for the volume collapse
transition.

\subsection{Global behavior}

Figure \ref{Ecorr} shows the correlation energy of the effective
LDA Hamiltonian Eq.~(\ref{Ham}), defined as the energy $E$
of Eq.~(\ref{Ham}) less the paramagnetic HF result $E_{\rm
PMHF}$ for the same Hamiltonian.  
Results for polarized HF, DMFT(H-I), and DMFT(QMC)
as obtained from Eqs.~(\ref{Eng1}) and (\ref{Eng2}) in the last two cases,
respectively, are compared in
this manner for an extended range of atomic volumes at five temperatures.
The polarized HF solutions assume
ferromagnetic spin order, and display both spin and orbital
polarization, with one band depressed below the Fermi level
and the other thirteen lying above.  These HF calculations
(dash-dot curves) are seen to give good energies at large volume
and low-temperature in comparison to the DMFT(QMC) (solid lines
with data points), as has been observed previously for the Anderson
Hamiltonian.\cite{HUSCROFT,PAIVA} Thus, the polarized Hartree-Fock
solution and other polarized static mean-field methods such as
orbitally polarized LDA,\cite{Eriksson90,Soderlind02,Svane97}
self-interaction-corrected LDA,\cite{Svane97,Svane94,Szotek94}
and LDA+U\cite{Sandalov95,Shick01} can be expected to give good
low-temperature total energies in the strong coupling limit.
As the atomic volume is reduced, the difference between polarized
and paramagnetic HF energy, $E_{\rm polHF}\!-\!E_{\rm PMHF}$,
becomes positive near $22$ \AA$^3$, and the HF solution has a
transition from the polarized to the paramagnetic solution, where
all fourteen bands have coalesced together above but slightly
overlapping the Fermi level.  The highest two temperatures in
Fig.~\ref{Ecorr} lie above the critical point for this transition,
and so there is no polarized HF solution.

\begin{figure}[tb]
 \includegraphics[width=3.2in]{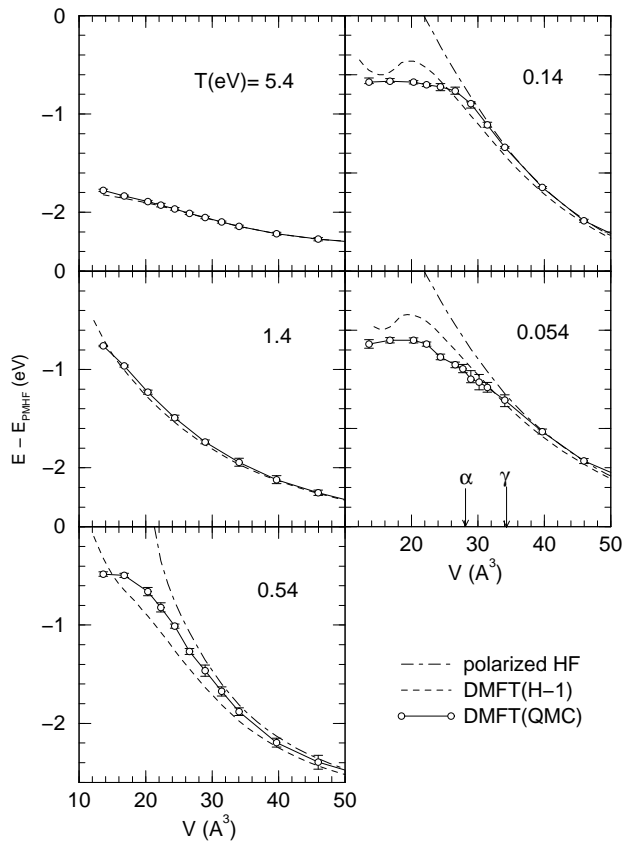}
\caption{Correlation energy, i.e., the difference between the
LDA+DMFT and the paramagnetic HF (PMHF) energy, as a function of
volume at five temperatures.  At large volumes, the LDA+DMFT(QMC)
energy agrees with the polarized HF and the Hubbard-I (H-I)
solutions.  But the LDA+DMFT(QMC) energy breaks away from the
polarized HF energy for decreasing volume, leading to a region
of negative curvature in the vicinity of the experimentally
observed $\alpha$-$\gamma$ transition (indicated by arrows)
at low temperature.
\label{Ecorr}}
\end{figure}

Turning to the Hubbard-I approximation, which becomes exact in
the atomic limit, it is no surprise that the DMFT(H-I) results
(short-dashed curves) should agree well with the DMFT(QMC) energies
at large volume for {\it all} temperatures.  This approximation
is also exact in the high temperature limit, as may be seen from
Fig.~\ref{Ecorr}, where there is also increasingly good agreement
at high temperature for all volumes considered here.  The agreement
between the two distinct DMFT calculations in these limits provides
a test of the reliability of both approaches used here.

A direct view of the temperature dependence is given in
Fig.~\ref{EvsT}a where the energy $E$ of Eq.~(\ref{Ham}) is plotted
versus $T$ for an atomic volume $V$ of $46\,$\AA$^3$.  At this
relatively large volume, the DMFT(QMC) and DMFT(H-I) results agree
closely and smoothly interpolate between the polarized HF energy at
low temperatures and the paramagnetic HF result at high temperature
(above about 15 eV, not shown).  There is no temperature-induced
transition in the DMFT results here, in contrast to the unphysical
transition from the paramagnetic to the polarized phase within
HF at $T\!\sim\!1$ eV.  This transition is a shortcoming of the
paramagnetic HF phase in which double-occupations of $f$-electrons
on the same Ce site can  not be avoided such that the paramagnetic
(interaction) energy is too high.
\begin{figure}[tb]
 \includegraphics[width=3.2in]{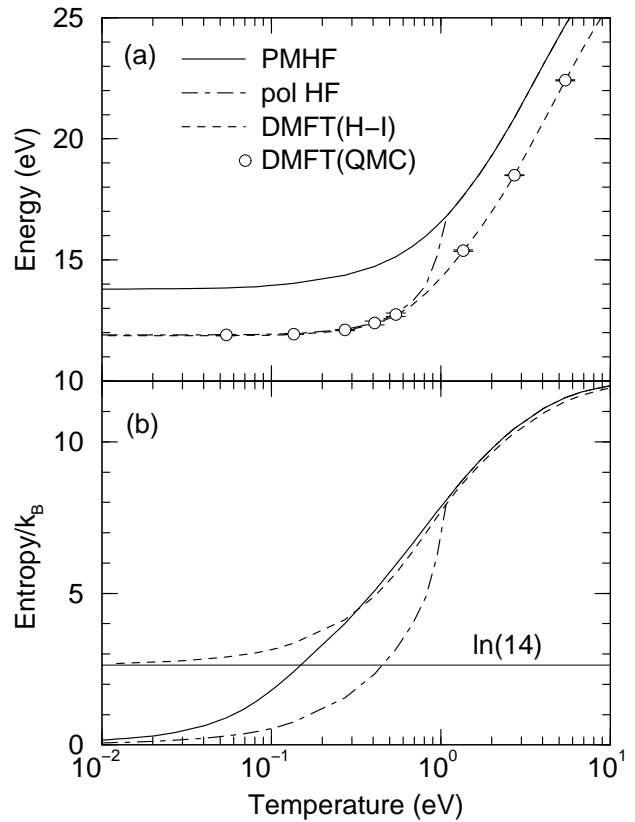}
\caption{
Energy (upper figure) and entropy (lower figure)
of the LDA+DMFT Hamiltonian Eq.~\protect(\ref{Ham})
vs.\ temperature at $V=46\,$\AA$^3$. 
At this relatively large volume, the DMFT(QMC) and DMFT(H-I) energies agree with
each other and, at lower temperatures, also with the polarized Hartree-Fock solution.
However, the entropy of the latter is completely wrong since 
the 14-fold degeneracy of the local magnetic moment is disregarded.
\label{EvsT}}
\end{figure}

Additional insight is provided by the corresponding entropy
in Fig.~\ref{EvsT}b.  The DMFT(H-1) entropy approaches $k_{\rm
B}\ln(14)$ at low temperature, which is effectively the degeneracy
of the Hund's rules magnetic moment $k_{\rm B}\ln(2J\!+\!1)$,
where without intra-atomic exchange and spin orbit interaction
we get the full 14-fold degeneracy of the $f$ level rather than
the proper 6-fold degeneracy for $J\!=\!5/2$.  At still lower
temperatures, crystal field effects are known to reduce the
entropy.\cite{Manley02}

Figure \ref{EvsT} illustrates two important aspects in which HF
and more rigorous techniques differ.  First, the HF transition at
about 1 eV corresponds to {\it simultaneous} moment formation and
magnetic ordering. In contrast, the two processes are distinct in
more rigorous treatments, with the moment formation occurring in
a continuous fashion at higher temperatures, culminating in the
low-$T$ plateau in Fig.~\ref{EvsT}b, with the onset of magnetic
order (if it occurs) coming at yet lower temperatures off the scale of the plot.
Second, polarized HF gives good {\it low}-$T$ energies at large
volumes because one of the Hund's rules multiplet states will
be a single Slater determinant.  However, its broken symmetry
mistreats the entropy at lower temperatures,  giving zero 
instead of, e.g., $k_{\rm B}\ln{(2J\!+\!1)}$ for $n_f\!=\!1$ in
the atomic limit, so that the finite-$T$ thermodynamics are
incorrect.

\subsection{Transition}

We now consider thermodynamic evidence for the $\alpha$--$\gamma$
transition in Ce.  While the QMC error bars restrict us from
making a quantitative prediction, we argue that the present
results do suggest the transition.  Evidence is already apparent
in Fig.~\ref{Ecorr}, where the DMFT(QMC) correlation energy is
seen to bend away from the polarized HF result as temperature
is lowered, leading to a region of negative curvature in the
vicinity of the observed transition (arrows).  As the other terms
($E_{\rm LDA}$ and $E_{\rm PMHF}-E_{\rm mLDA}$) contributing to
the total energy Eq.~(\ref{Etoteqn}) all have positive curvature
throughout the range considered in this work, this correlation
contribution is then the only candidate to create a region of
negative bulk modulus in the low-temperature total energy, i.e.,
a thermodynamic instability, and thence a first order phase
transition given by the Maxwell common tangent.

Figure \ref{Etotfig} shows total energies Eq.~(\ref{Etoteqn})
for the DMFT(QMC) and polarized HF methods at the three lowest
temperatures of Fig.~\ref{Ecorr}.  The region of negative curvature
just noted in the correlation energy is seen to cause a substantial
depression of the DMFT(QMC) total energies (solid curves with
symbols) away from the polarized HF results (dashed curves) below
35 \AA$^3$, which is most pronounced at the lowest temperature,
$T\!=\!0.054\,$eV.  The consequent shallowness in the DMFT(QMC)
curve at this temperature persists over the observed range of
the two-phase region (arrows), although statistical uncertainties
preclude any claim of seeing negative curvature.  The slope is also
consistent with a $-0.6$ GPa pressure (long-dashed line), which is
the extrapolated transition pressure at $T\!=\!0$.\cite{JOHANSSON2}
We suggest in fact that these $T\!=\!0.054\,$eV (632 K) total
electronic energies are already close to the low-$T$ limit.  Both
the DMFT(H-I) and HF energies at this temperature differ by less
than 0.006 eV from corresponding results at half this temperature,
throughout the volume range in Fig.~\ref{Etotfig}.  Our DMFT(QMC)
calculations at $T\!=\!0.027\,$eV (316 K) are also consistent
with this conclusion, as discussed in Sec.~\ref{validation},
\begin{figure}[tb]
 \includegraphics[width=3.2in]{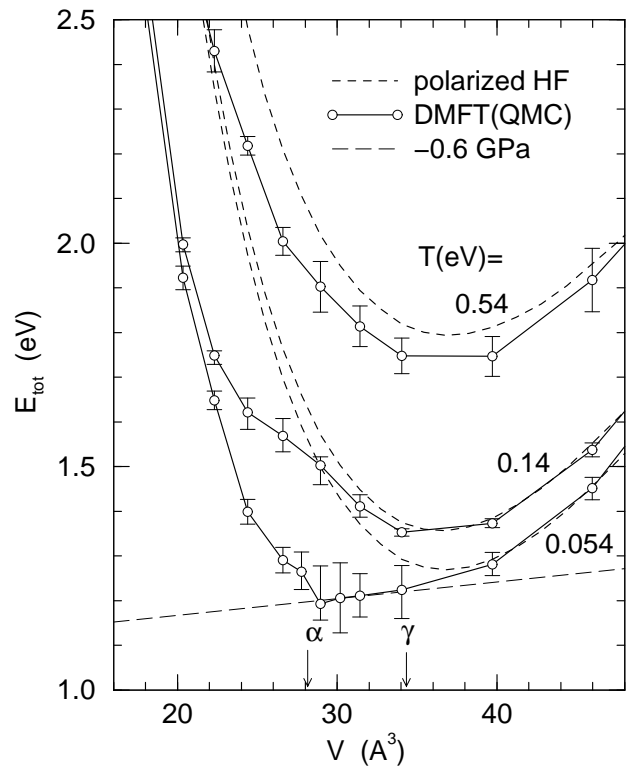}
\caption{Total LDA+DMFT(QMC) and polarized HF energy as a function
of volume at three temperatures. While the  polarized HF energy
has one pronounced minimum in the $\gamma$-Ce phase, the negative
curvature of the correlation energy of Fig.~\ref{Ecorr} results in
the development of a side structure ($T\!=\!0.14\,$eV), and finally a
shallowness ($T\!=\!0.054\,$eV), which is consistent with the observed
$\alpha$-$\gamma$ transition (arrows) within our error bars.
These results are also consistent with the experimental pressure
given by the negative slope of the dashed line.
\label{Etotfig}}
\end{figure}

That the electronic contribution to the total energy might be
close to its low temperature limit below about $600\,$K is also
consistent with the analysis of the $\alpha$--$\gamma$ transition
by Johansson {\it et al.},\cite{JOHANSSON2} who attribute the
temperature dependence of the transition pressure primarily to the
difference in entropy, which is zero and $k_{\rm B}\ln{(2J\!+\!1)}$
for the $\alpha$ and $\gamma$ phases, respectively.  That is,
for temperatures larger than both the Kondo temperature and the
crystal-field splitting\cite{Manley02} in the $\gamma$ phase,
yet still fairly low (say $200$--$600\,$K), the temperature
dependence of the $\gamma$-phase free energy may be dominated
by the linear term $-k_{\rm B}\ln{(2J\!+\!1)}T$ arising from a
plateau such as in Fig.~\ref{EvsT}b, while presumably the total
energies (both $\alpha$ and $\gamma$) are closer to the low-$T$
limit due to their faster $T^2$ dependence.

We have calculated both the DMFT(QMC) specific heat $C(V,T)$
and entropy $S(V,T)$ for the effective Ce LDA Hamiltonian
Eq.~(\ref{Ham}).  We first calculated DMFT(H-I) energies
Eq.~(\ref{Eng1}) on a logarithmic temperature grid up to the
high-$T$ limit ($\sim \!10^3$ eV) where the entropy is known to
be 12.057$\,k_{\rm B}$.  As noted earlier, the DMFT(H-I) method
is correct at high temperatures, and indeed the DMFT(QMC) energies
obtained via Eq.~(\ref{Eng2}) closely approach the H-I results as
$T$ is increased, e.g., lying above by only 0.024 and then 0.004 eV
at $T=5.4$ and 13.6 eV, respectively, for $V\!=\!16.8\,$\AA$^3$.
We therefore fit the difference between the QMC and H-I energies
at eight temperatures from 0.054 to 5.4 eV to the form $a+\sum_n
b_n/(1+n\Delta/T^2)$, $n\!=\!1$--3, which has a $T^2$ behavior
at low temperatures, and is benign at high temperatures.  These
smoothed and interpolated differences were added to the DMFT(H-I)
energies to create a fine grid of ``DMFT(QMC)'' energies from
which $C(V,T)\!=\!\partial E(V,T)/\partial T|_V$ was calculated
by numerical differentiation, and $S(V,T)$ by integration down
from the high-$T$ limit according to Eq.~(\ref{Seqn}).  Note that
while the finite nature of Eq.~(\ref{Ham}) is unphysical at
very high temperatures, these results are nonetheless entirely
meaningful at more modest temperatures where the omitted core and
higher-lying valence states will be frozen out, e.g., below $\sim
3$ eV near the $\alpha$--$\gamma$ transition, given a spectrum of
Eq.~(\ref{Ham}) that extends to nearly $30\,$eV above the Fermi
level in that volume range.

The challenging need for accurate energy derivatives, as well as
the sensitivity of Eq.~(\ref{Seqn}) to the lowest temperatures
given the $1/T$ factor, requires a stringent convergence
criterion for the kinetic-energy Matsubara sums. Otherwise
we observe unphysical negative low-$T$ limits of the entropy
for $V\!<\!25\,$\AA$^3$.  We have also constrained the fits to
smooth out the value of this low-$T$ limit as a function of volume
over this same range.  In all cases it is to be emphasized that
the fits give excellent representation of features in $E_{\rm
DMFT(QMC)}(T)-E_{\rm DMFT(H-I)}(T)$, ranging in size from 0.1 to
0.24 eV upon decreasing the volume from $V\!=\!35$ to 25 \AA$^3$,
and are well within the $\pm 0.03$ eV error bars in the data.
The same fits were used to obtain both $C(V,T)$ and $S(V,T)$.

\begin{figure}[tb]
 \includegraphics[width=3.2in]{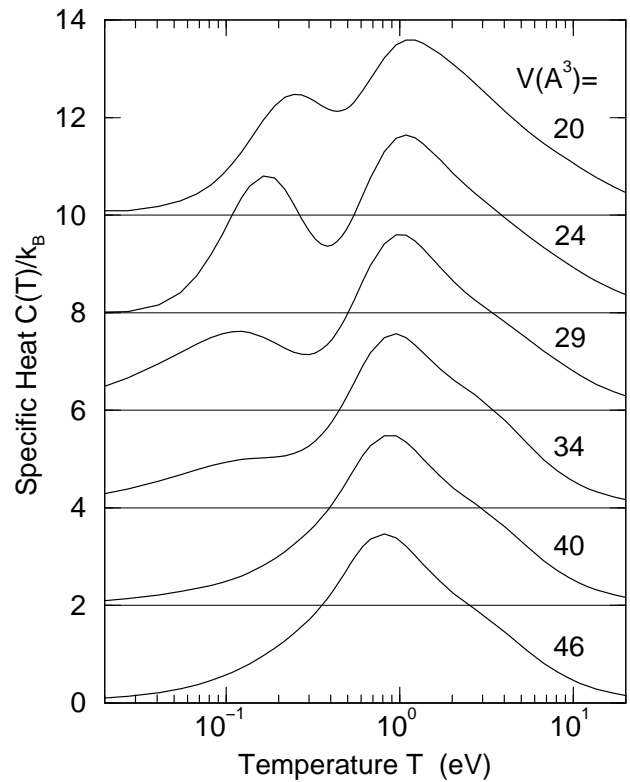}
\caption{Specific heat as a function of temperature for different
volumes (off-set as indicated). At smaller volumes, an additional
low-energy peak develops, coinciding with the formation of an
Abrikosov-Suhl resonance (see Fig.~\ref{fSpectrum} below).
\label{CvsT}}
\end{figure}

The temperature dependence of the DMFT(QMC) specific heat is
shown in Fig.~\ref{CvsT} at six volumes.  The most significant
feature is the appearance of the low temperature peak in the
range $T\!=\!0.1$--$0.2\,$eV, which coincides precisely with
growth of the quasiparticle peak or Abriksov-Suhl resonance at
the Fermi level in the $4f$ spectra, as will be seen in the
next section.  Analogous behavior has been discussed for the
one-band Hubbard model.\cite{DMFT2} The low temperature peak in the
specific heat is just barely discernible at the $\gamma$-phase
volume of $34\,$\AA$^3$ in Fig.~\ref{CvsT}, has become rather
prominent by $29\,$\AA$^3$, which is slightly larger than the
$\alpha$-phase volume, and then continues to broaden and shift
to higher temperatures as volume is further reduced.  The broad
peak near $1\,$eV which appears at all volumes is due both to
the $4f$ charge fluctuations, and also to $spd$-valence to $4f$
excitations, given that $n_f$ increases by $\sim\!20$\% on raising
the temperature to $1.4\,$eV.  Note also in regard to the charge
fluctuations that the peak in $C(T)$ should occur at significantly
smaller $T$ than the Coulomb repulsion $U_f\!\sim\!6\,$eV, as may
be seen in the case of the half-filled one-band Hubbard model.
Here, the specific heat peak occurs at $T\!=\!0.208\,U$ in the
absence of hopping $t\!=\!0$, and the location of the peak is
also depressed by the band width.\cite{DMFT2}
\begin{figure}[tb]
 \includegraphics[width=3.2in]{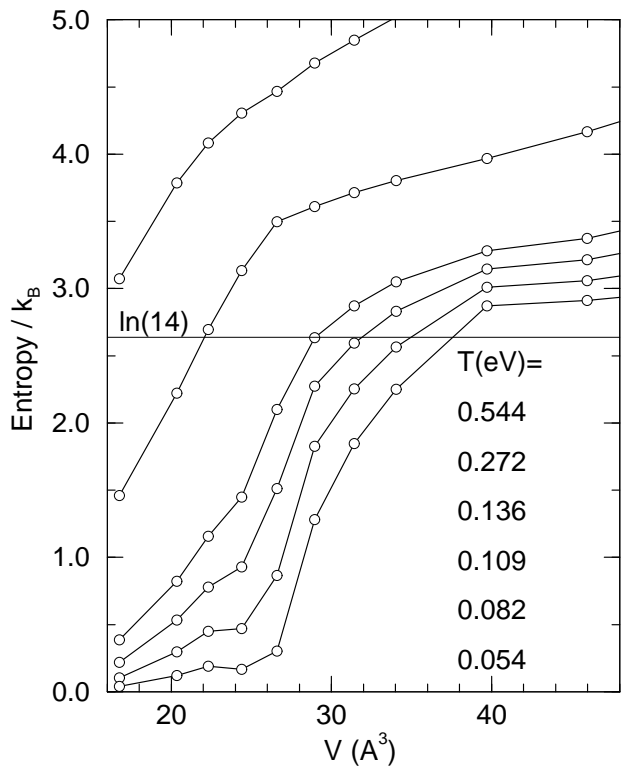}
\caption{Entropy as a function of volumes for different temperatures.
In the vicinity of the $\alpha$-$\gamma$ transition (28.2-34.3 \AA$^3$),
the entropy increases rapidly.
\label{SvsV}}
\end{figure}

The volume dependence of our DMFT(QMC) entropy is shown in
Fig.~\ref{SvsV} for six temperatures. The rapid increase
in the entropy over the $\alpha$--$\gamma$ transition
($28.2$--$34.4\,$\AA$^3$) is due precisely to the low temperature
peak peak in $C(T)$, which contributes to the entropy via its
weighted area $\int dT C(T)/T$.  Thus, at large volumes where
the $4f$ spectral weight is Hubbard-split with no contribution
at the Fermi level, the low-$T$ entropy is pinned at $k_{\rm
B}\ln(2J\!+\!1)$ (ignoring effects of crystal field at yet
lower $T$).  Then, as the volume is reduced, the quasiparticle
peak begins to grow at the Fermi level, the weighted area of its
associated heat capacity peak reduces the low-$T$ entropy below
$k_{\rm B}\ln(2J\!+\!1)$ via Eq.~(\ref{Seqn}).  The physical
interpretation is of course that the degeneracy associated with
the $2J\!+\!1$ directions of the Hund's rules moment disappears as
this moment is either screened or collapses on reducing the volume.

For completeness, we conclude this section by providing
the free energy $F=E_{tot}\!-\!ST$ in Fig.~\ref{Ftotfig},
although the uncertain errors in the entropy, and the fact
that the large-$V$, low-$T$ value is 50\% too large [taking
into account the spin-orbit coupling will give
$k_{\rm B}\ln(6)$ instead of $k_{\rm B}\ln(14)$].
Given that the electronic total energy $E_{tot}$ is near its
low-$T$ limit by $T\!=\!0.054$ eV, we consider that curve as
``$T\!=\!0$'', and then include it again as $F=E_{tot}\!-\!ST$
for $T\!=\!0.054$ eV.  The error bars on all curves are just
from the energy.  The slopes of the two straight lines give the
experimental transition pressures at $T\!=\!0$ and $0.054$ eV,
and arrows mark the observed boundaries of the $\alpha$--$\gamma$
transition at room temperature.  The essential conclusion of
Fig.~\ref{Ftotfig} is that these results are consistent with
experiment, though stronger claims are precluded by the statistical
uncertainties.  Nonetheless, the results of this section which we
find compelling are the way in which $E_{tot}(V)$ systematically
develops a shallowness in the vicinity of the $\alpha$--$\gamma$
transition as temperature is lowered, and the structure in the
specific heat and entropy.

\begin{figure}[tb]
 \includegraphics[width=3.2in]{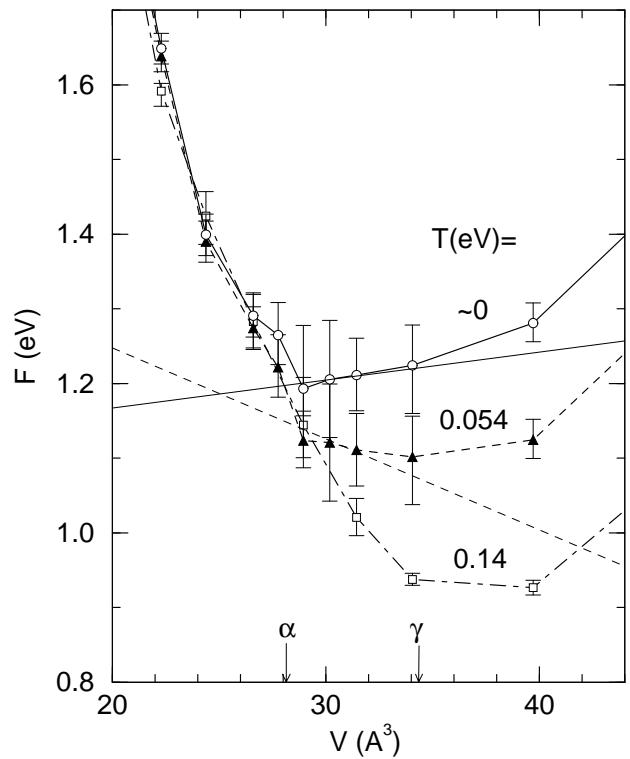}
\caption{Free energy as a function of volume at three temperatures,
compared to lines whose negative slopes give the experimental
$\alpha$-$\gamma$ transition pressures at $T\!=\!0$ (solid line)
and 0.054$\,$eV (dashed line). Given the statistical uncertainties,
the results are consistent with experiment and show that a shift
of the $\alpha$-$\gamma$ transition volumes is primarily due to
the entropy.
\label{Ftotfig}}
\end{figure}
\section{Spectra}

In this section, we discuss the spectral changes through the
$\alpha$-$\gamma$ transition. To obtain the  physical spectrum
$A(\omega)=-\frac{1}{\pi} {\rm Im} G(\omega)$, one has to
analytically continue the QMC data from the imaginary time
(Matsubara frequency) representation to real frequencies $\omega$:
\begin{equation}
G(\tau)=\int_{-\infty}^{\infty} {\rm d} \omega \; \frac{e^{\tau
(\mu-\omega)}}{1+e^{\beta (\mu-\omega)}} A(\omega).
\label{AnaCont}
\end{equation}
As one can see in Eq.~(\ref{AnaCont}), the values of $A(\omega)$
at large (positive or negative) frequencies affect $G(\tau)$
only weakly because the integral kernel is exponentially small
in this regime.  To deal with this ill-conditioned problem which
is particularly cumbersome in the  presence of the statistical
QMC error, we employ the maximum entropy method.\cite{MEM}
When interpreting the results later on, we have to keep in mind,
however, that there is a significant error at larger frequencies
which tends to smear out fine features  such that, e.g.,
inner structures of Hubbard bands are not necessarily resolved.
In  section \ref{SpectraChange}, we present the spectra of the
f- and valence-electrons of fcc Ce as a function of volume and
discuss the changes at the $\alpha$-$\gamma$ transition. The
spectra obtained are compared to photoemission and Bremsstrahlung
experiments in Section~\ref{SpectraExp}.

\subsection{Change of the spectra at the $\alpha$-$\gamma$ transition}
\label{SpectraChange}

In Section \ref{TD} we noted a region of negative curvature in the
correlation energy at volumes consistent with the experimental
$\alpha$- and $\gamma$-volumes, leading to a shallowness in
the total energy and suggesting a first order phase transition
at lower temperatures.  To further elucidate the nature of
the ongoing changes, we study the evolution of the $f$-electron
spectrum as a function of volume for fcc Ce at $T\!=\!0.054\,$eV
(632 K) in Fig.~\ref{fSpectrum}. This temperature is close to the
critical endpoint ($T\!=\!600 \pm 50\,$K) at which the first order
$\alpha$-$\gamma$ transition disappears experimentally.\cite{Ce}
From the continuous evolution of the energy versus volume curves,
we expect, however, similar changes above the critical endpoint,
which are not yet strong enough to cause a first order phase
transition.  At a very small volume, $V\!=\!20\,$\AA$^3$, most
of the spectral weight is seen to be in a big quasi particle
peak or Abrikosov-Suhl resonance at the Fermi energy, but some
spectral weight has already been transferred to side structures
which would be interpreted as upper and lower Hubbard bands in a
Hubbard model.  Moving closer to the $\alpha$-$\gamma$ transition
(between $28.2$ and $34.4\,$\AA$^3$ at room temperature), the
$\alpha$-Ce-like spectra at $V=29\,$\AA$^3$ shows this three peak
structure to become more pronounced with a sharp Abrikosov-Suhl
resonance and well-separated Hubbard bands.  The spectral
weight of the Abrikosov-Suhl resonance is further reduced and
smeared out when going to the $\gamma$-phase ($V=34\,$\AA$^3$)
and finally disappears at large volumes ($V=46\,$\AA$^3$), at
least at $T=0.054\,$eV.
\begin{figure}[tb]
 \includegraphics[width=3.2in]{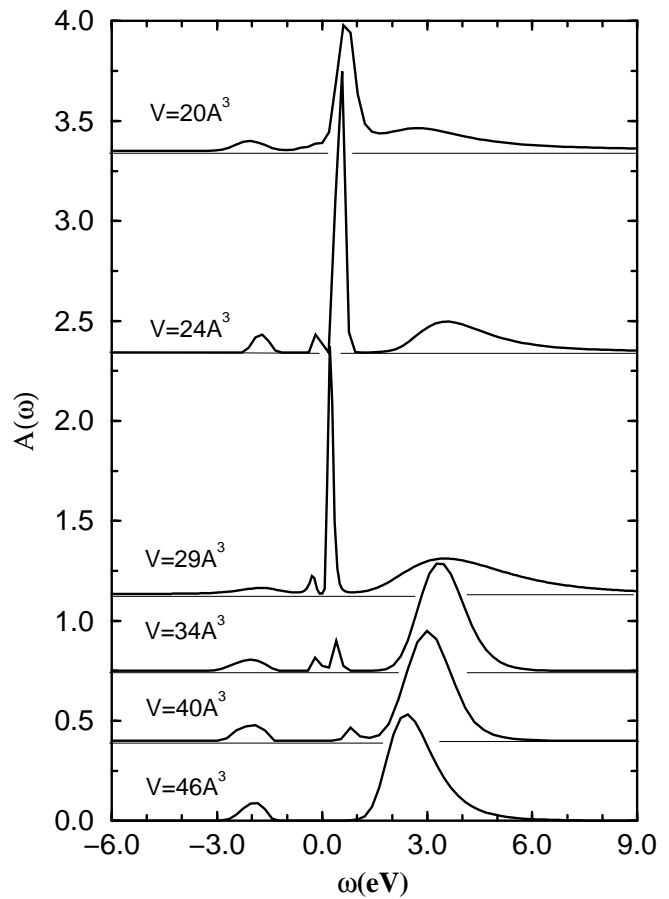}
\caption{Evolution of the $4f$ electron spectrum with volume
at $T=632\,$K;  off-set as indicated. When
going from small to large volume, the
weight of the central Abrikosov-Suhl resonance decreases 
and practically fades away  at the  $\alpha$-$\gamma$ transition
from $V=29$  to  $34$\AA$^3$ . The residual weight
around the Fermi energy at   $V=34$\AA$^3$  indicates
a smeared out Abrikosov-Suhl resonance as is to be expected if
 the Kondo temperature of $\gamma$-Ce is below $T=632\,$K.
\label{fSpectrum}}
\end{figure}

Altogether,  we observe, as a function of volume, the crossover
from a structure which differs only slightly from a one-peak
structure, to a three peak structure, and finally  to a two peak
structure.  The physical interpretation is that the $f$-electrons are
 somewhat correlated at low volumes, where the large
quasiparticle peak above the Fermi energy  
resembles (to a first approximation) the one-peak structure
of the uncorrelated one-particle theory or the LDA.  At larger
volumes, the system is highly correlated, there is a magnetic
moment imposed by the electrons in the lower Hubbard band, but
the $f$-electrons at the Fermi energy are still itinerant. Finally
at the largest volumes, the $f$-electrons are localized and the
local magnetic moment is fully developed. Here, the most dramatic
change of the weight of the quasiparticle peak coincides with the
observed region of negative curvature in the correlation energy.
We thus conclude that the drastic reduction of the weight
of the quasiparticle peak is related to the energetic changes
in the correlation and total energies which are consistent with
the first order $\alpha$-$\gamma$ transition.  

These features and also the three peak (Kondo-like) structure of
the $\alpha$ and $\gamma$ phases agree with the Kondo volume
collapse scenario.\cite{KVC} On the other hand, many-body
calculations show that the behavior of the Anderson and Hubbard
models --- paradigms for the Kondo volume collapse\cite{KVC} and
Mott transition\cite{JOHANSSON} scenarios, respectively --- are
remarkably similar in regard to their spectra and other properties
at finite temperatures.\cite{Held00ab} One important difference,
however, is the absence of spectral weight at the Fermi level in
the ``large volume'' phase of the Mott-Hubbard transition, as for
example in V$_2$O$_3$,\cite{Held01a} in contrast to the reduced
but still extant spectral weight in our $\gamma$-Ce results
and the experiment,\cite{Liu92}
which is a more Kondo-like feature.

In Fig.~\ref{fTSpectrum}, we compare the $4f$-spectrum in the
$\alpha$  and $\gamma$ phases to results at higher temperatures
($T\!=\!0.14\,$eV) from Ref.~\onlinecite{Held01b}. Most
notably, the Abrikosov-Suhl resonance in the $\alpha$-phase
($V=29\,$\AA$^3$) becomes much sharper when going from
$T\!=\!0.14\,$eV to $0.054\,$eV. The reason for this is that
 the Abrikosov-Suhl 
resonance is smeared out  thermally at 
$T\!=\!0.14\,$eV ($1580$ K) since this temperature
 is comparable
to the Kondo temperature, which we estimate
to be  $0.18\,$eV ($ 2100\,$K) from  the full width at half 
maximum [LDA+DMFT(NCA) calculations yield 1000 K, see 
Ref.~\onlinecite{Zoelfl01}].
This Kondo temperature is only a crude estimate which 
might also be somewhat reduced if the spin orbit coupling, 
which splits off states from the Fermi energy, is taken into account.
Nonetheless, it reasonably agrees with experimental
estimates of $T_{\rm K}\!=\!945\,$K and $1800$--$2000\,$K for
the Kondo temperature obtained from electronic\cite{Liu92} and
high-energy neutron spectroscopy \cite{Murani93}, respectively.  In
contrast, the peak in the $\gamma$ phase remains smeared out such
that one would assume a Kondo temperature lower than $0.054\,$eV
($632$ K); the experimental estimates are $T_{\rm K}=95\,$K
(Ref.~\onlinecite{Liu92}) and $60\,$K (Ref.~\onlinecite{Murani93}).
The changes in the rest of the spectrum are much less dramatic.
The position of the upper Hubbard band is fixed while the lower
Hubbard band, which has a very small spectral weight, moves closer
to the Fermi energy upon decreasing the temperature.
\begin{figure}[tb]
\includegraphics[width=3.2in]{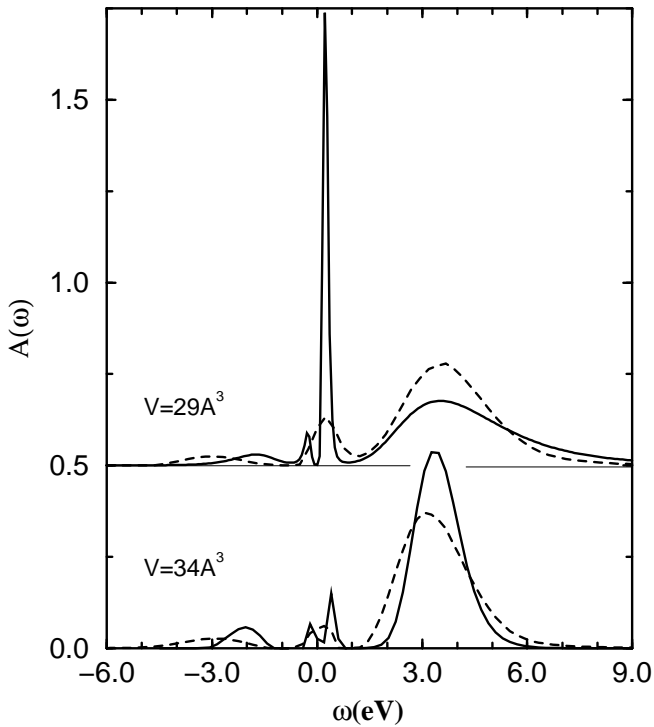}
\caption{$4f$ electron spectrum for $\alpha$-Ce ($V=29$\AA$^3$) and $\gamma$-Ce ($V=34$\AA$^3$)
at two temperatures ($T=632\,$K: solid line; $T=1580\,$K: dashed line).
The Abrikosov-Suhl resonance of  $\alpha$-Ce is smeared out when increasing the
temperature from $T=632\,$ to $1580\,$K, indicating that the Kondo temperature is in between.
\label{fTSpectrum}}
\end{figure}

While the $f$-electrons undergo a transition from itinerant
character at low volumes with a quasiparticle resonance at the
Fermi energy, to localized character at larger volumes without
such a resonance, the ($spd$) valence electrons remain metallic
at all volumes.  This can be seen in Fig.~\ref{vSpectrum}, which
shows the valence spectral function $A(\omega)$ averaged over the
$spd$-orbitals. It is finite at the Fermi energy for all volumes,
such that Ce is always a metal.  The biggest change in the spectrum
is the decreasing valence bandwidth when increasing the volume,
which is simply due to the reduced overlap of the valence orbitals
as inter-atomic distances increase.  The effect of electronic
correlations is less  obvious.  But, one can note a dip in the
valence spectrum in the vicinity of the Fermi energy which is to
be expected to coincide with the Abrikosov-Suhl resonance in the
$f$-spectrum. This dip is most pronounced at lower volumes where
the $f$-electron Abrikosov-Suhl resonance has most spectral weight.
\begin{figure}[tb]
\includegraphics[width=3.2in]{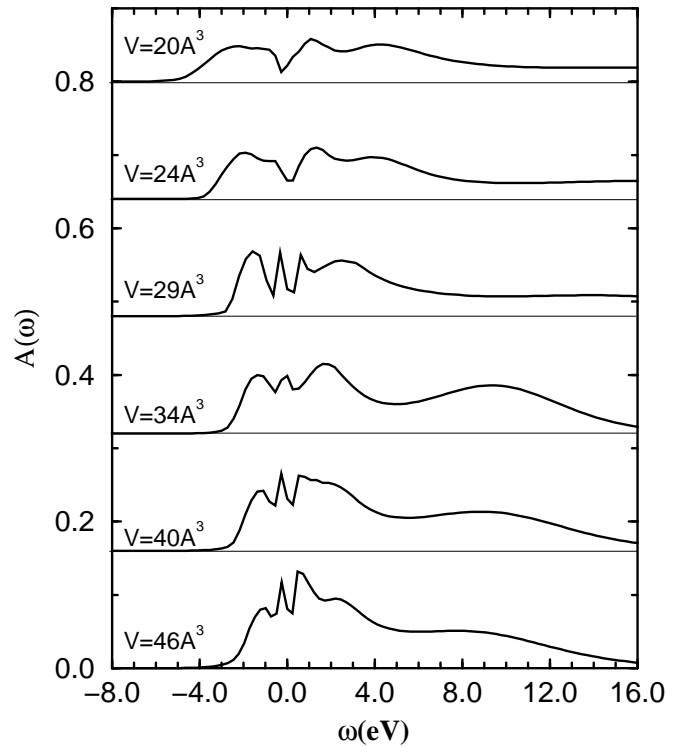}
\caption{Evolution of the $spd$ electron spectrum with volume
at $T=632\,$K; off-set as indicated.
Note the wider energy window in comparison to Figs.~\protect\ref{fSpectrum}
and ~\protect\ref{fTSpectrum}.
The main effect to be seen  is the decrease of the bandwidth upon increasing the volume.
\label{vSpectrum}}
\end{figure}

\subsection{Comparison to experiment}
\label{SpectraExp}

The LDA+DMFT(QMC) calculation of fcc Ce suggests a volume collapse
approximately at the experimental volumes.  To further test
whether this theory actually describes fcc Ce, we now  compare our
$\alpha$- and $\gamma$-Ce spectra with photoemission spectroscopy
(PES)\cite{Wieliczka84} and Bremsstrahlung isochromatic
spectroscopy (BIS).\cite{Wuilloud83} To this end, we combined
the $f$ and $spd$ spectra of Section \ref{SpectraChange} with
areas normalized to 14 and 18, respectively, to yield the full
$spdf$ density of states, and smoothed it with the experimental
resolution of approximately $0.4\,$eV.

The comparison is shown in Fig.~\ref{totSpectrum} for $\alpha$
and $\gamma$ Ce. Although there are no free parameters in our
LDA+DMFT(QMC) results,\cite{NoteSpectra} the agreement between
theory and experiment is very good.  Particularly good is the
agreement of the spectrum around the Fermi energy for both $\alpha$
{\it and} $\gamma$ Ce;  this part of the spectrum consists of
the Abrikosov-Suhl resonance of the $f$-electron spectrum and the
valence spectrum.  Also the position of the upper and lower Hubbard
bands and the relative weight of these peaks and the Abrikosov-Suhl
resonance is correctly predicted by the theory.  Less good is the
agreement with respect to the width of the upper Hubbard band which
is too narrow in our theory; the experimental upper Hubbard bands
extend to energies 1--2$\,$eV higher than our theory.  
As has been argued in Ref.~\onlinecite{Zoelfl01}, this can be
understood by the Hund's rules exchange coupling which has not been taken
into account in our calculation. We justified this by noting that
the exchange coupling
 is only effective if there are more than two electrons on one
Ce site which happens only rarely. However, the excited states of
the upper Hubbard-band correspond to just such double occupied
states. For these, the Hund's rules coupling becomes important
and will split the upper Hubbard band into multiplets. With this
shortcoming resolved, the comparison to the experimental spectrum
suggests that our LDA+DMFT(QMC) calculation describes $\alpha$
and $\gamma$ Ce very well.
\begin{figure}[tb]
 \includegraphics[width=3.2in]{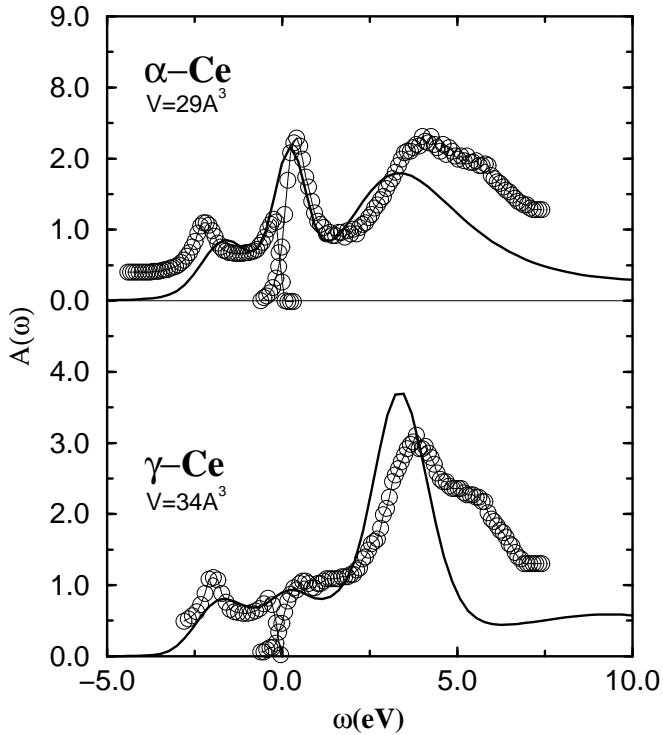}
\caption{
Comparison of the LDA+DMFT(QMC) spectra with experiment (circles)\cite{Liu92}.
 Although there
are no free parameters in the calculated spectrum,
the agreement is very good, in particular at the Fermi energy($\omega=0$).
The additional structure in the upper Hubbard band which
is seen in the experiment is likely due to the exchange
interaction which was neglected in our calculation.
\label{totSpectrum}}
\end{figure}

The $\alpha$ and $\gamma$ spectra of previous LDA+DMFT(NCA)
calculations by Z\"olfl {\it et al.} \cite{Zoelfl01} are
considerably different from ours and the experimental spectra,
in particular the weight of the upper Hubbard bands was much
higher in Ref.~\onlinecite{Zoelfl01}.   The temperature of
Ref.~\onlinecite{Zoelfl01} is very close to ours ($T\!=\!580\,$K)
and also the $4f$-electron Coulomb interaction value $U_f$
is comparable, at least for the $\gamma$ phase; Z\"olfl
{\it et al.} employed a fixed value of $U_f\!=\!6\,$eV whereas the
constrained LDA values in our calculations are $U_f\!=\!5.72\,$eV
and $5.98\,$eV for $\alpha$ and $\gamma$ Ce, respectively. In
view of this we tend to explain the differences, at least
for  $\gamma$-Ce, by the different
method employed to solve the DMFT equations, in particular,
since the non-crossing approximation (NCA) is a resolvent
perturbation theory for strong coupling.

\section{Local moment and $4f$ occupations}

Important additional information about the $\alpha$-$\gamma$
transition and the effects of electron correlation in Ce are
contained in the number of $4f$ electrons per site $n_f$,
the double occupation $d$, and quantities derived from these
such as the fraction of sites $w(f^n)$ with $n\!=\!0$,$1$,$2$
$f$ electrons, and the local magnetic moment.  These parameters
can discriminate between the various models, as for example the
promotional model \cite {PM} assumes a considerable change in the
number of $4f$ electrons at the $\alpha$-$\gamma$ transition,
in contrast to the Kondo Volume collapse \cite {KVC} and Mott
transition\cite{JOHANSSON} scenarios which do not. The latter two
on the other hand distinguish themselves by assuming a small and
large change of the magnetic moment, respectively.

Figure \ref{nffig} gives $n_f$ as a function of volume at four
temperatures.\cite{nfmeas} The lowest curve at $T\!=\!0.054\,$eV
(632 K) is already very close to the low-$T$ limit, as our
results at half this temperature are the same to within generally
$0.004$, or at most $0.01$ electrons per site. There are two
main tendencies: With decreasing $V$,  $n_f$ increases due to
the upward motion of the $6s,p$ levels relative to the $4f$
level under compression; it also increases with $T$ due to the
thermal occupation of the large $4f$ density of states lying
above the Fermi level.  Superimposed on this behavior is, at
low temperatures, an abrupt reduction of $n_f$ in the observed
two-phase region (marked) as volume is reduced, an anomaly
which is annealed away by $T\!=\!0.5\,$eV similar to the case
of the total energy.  This effect leads to a number of $4f$
electrons close to one, ruling out the promotional model \cite
{PM} and suggesting Kondo physics given the sharp quasiparticle
peak seen in the previous section.  Quantitatively, we get a 4\%
reduction in $n_f$ across the two-phase region from $1.035\pm0.017$
to $0.993\pm 0.010$ at $T\!\leq0.054\,$eV.  Similar behavior
is seen in the 10\% drop from $1.014$ to $0.908$ of Z\"olfl
{\it et al.}\cite{Zoelfl01} in their LDA+DMFT(NCA)calculations,
and the 11\% reduction from $0.971\pm 0.006$ to $0.861\pm0.015$
electrons/site of Liu and coworkers,\cite{Liu92} who fitted a
single impurity Anderson model to the  experimental $4f$ spectrum.
The reason for the drop in $n_f$ is a systematic increase in
the double occupation $d$ under compression.  Since $d$ is the
potential energy divided by $U_f$, the energy cost associated
with increasing $d$ can be ameliorated by reducing $n_f$.

\begin{figure}[tb]
 \includegraphics[width=3.2in]{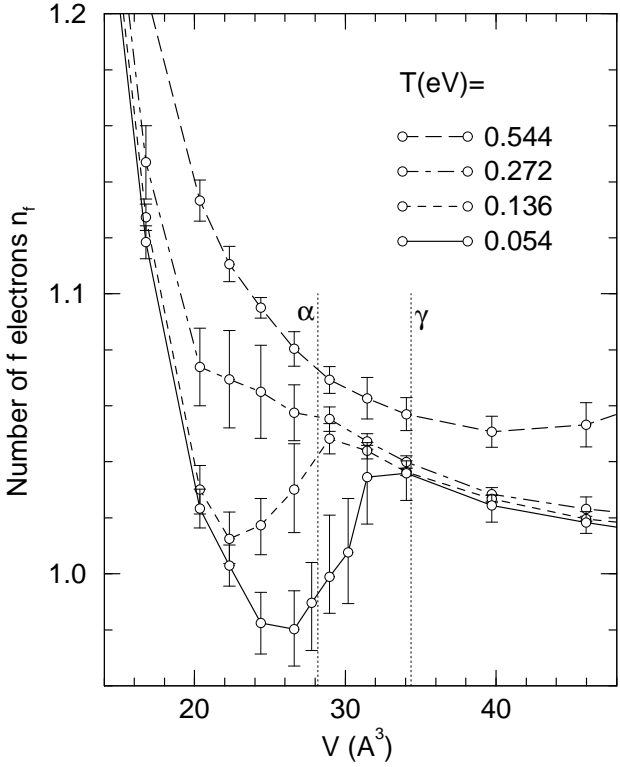}
\caption{
Number of $4f$ electrons $n_f$ vs.\ volume at four temperatures.
At low temperatures and in the vicinity
of the $\alpha$-$\gamma$ transition, $n_f$ is very close to one. 
\label{nffig}}
\end{figure}

Since there is little chance of more than doubly-occupied sites
in Ce at low temperature, $n_f$ and $d$ provide sufficient
information to obtain the fractions of sites with various integral
$f^n$ occupations.
\begin{eqnarray}
w(f^0) &=& 1-n_f+d
\nonumber \\
w(f^1) &=& n_f-2d
\nonumber \\
w(f^2) &=& d
\label{fneqn}
\end{eqnarray}
Fig.~\ref{wnfig} shows our DMFT(QMC) results for these weights at
$T\!=\!0.054\,$eV, which are also close to the low-temperature
limit.  At large volume one sees that each site nearly always
has one $f$ electron, and that empty or doubly occupied sites are
rare, as would be expected for $n_f\!\sim\! 1$ in the absence of
significant hybridization to move these electrons to either $f$
or $v$ ($spd$ valence) states on neighboring sites.  For the
$f$ electrons to begin to move around from one site to another
in any independent fashion under the influence of larger $ff$
hybridization, or for there to be virtual charge fluctuations
of the form $f^1v^3\rightarrow f^0v^4$ and $f^1v^3\rightarrow
f^2v^2$ due to increased $fv$ hybridization, it is clear in each
case that both empty and doubly occupied sites must become more
common at the expense of singly occupied sites if the volume is
reduced, as evident in Fig.~\ref{wnfig}.  Note that these changes
are especially dramatic over the experimental two phase region
(marked).
\begin{figure}[tb]
 \includegraphics[width=3.2in]{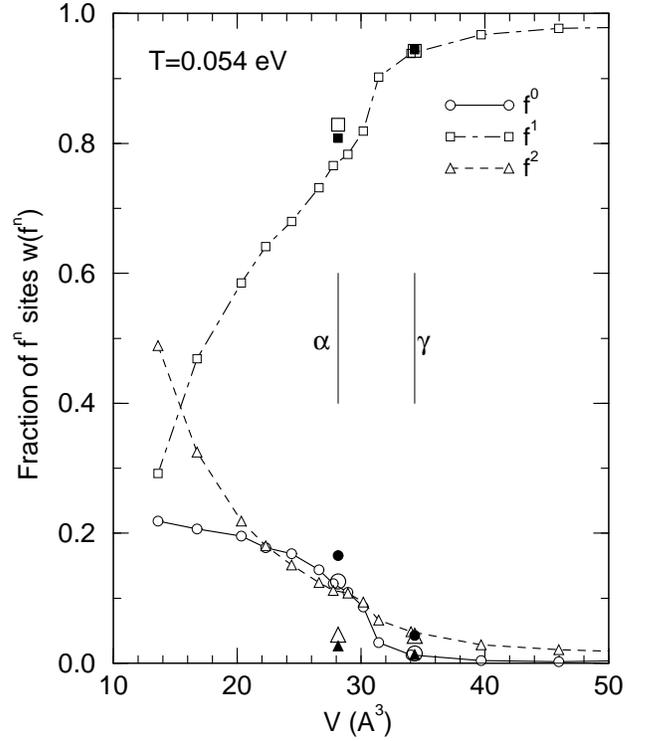}
\caption{Fraction of empty ($f^0$), singly ($f^1$), and
doubly occupied sites ($f^2$) vs.\ volume as calculated by
LDA+DMFT (QMC) (open symbols with lines) at $T=0.054\,$eV
in comparison to LDA+DMFT(NCA) (large open symbols)
\protect\cite{Zoelfl01} and impurity Anderson model results
(filled symbols). \protect\cite{Liu92} While the DMFT results
agree very well for the $\gamma$ phase, there are significant
differences in the $\alpha$ phase as discussed in the text.
\label{wnfig}}
\end{figure}

The filled symbols in Fig.~\ref{wnfig} show the impurity Anderson
model results of Liu {\it et al.}\cite{Liu92} at the observed
$\alpha$- and $\gamma$-Ce volumes; the large open symbols, the
DMFT(NCA) results of Z\"olfl and coworkers.\cite{Zoelfl01} Our
DMFT(QMC) values are $w(f^0), w(f^1), w(f^2)= 0.013\pm0.019$
($0.118\pm0.025$), $0.939\pm0.028$ ($0.771\pm0.033$), and
$0.048\pm0.009$ ($0.111\pm0.008$) for the $\gamma$ ($\alpha$)
volumes, respectively.  The two DMFT calculations agree well
within these uncertainties for all three populations $w(f^n)$
at the larger $\gamma$-phase volume, and also with the impurity
Anderson model value for $w(f^1)$; although for the two small
populations, they obtain $w(f^0)\!<\!w(f^2)$ in reverse order to
the values of Liu and coworkers.\cite{Liu92} The most significant
difference at the $\alpha$-Ce volume is the rather larger
double occupancy, $d\!=\!w(f^2)=0.111\pm 0.008$, obtained by our
DMFT(QMC) calculations in comparison to smaller values 0.044 and
0.026 obtained by the the DMFT(NCA) and impurity Anderson model,
respectively.  Temperature is unlikely to be a factor here, as we
obtain $n_f$ and $d$ unchanged within our error bars at half the
temperature of the DMFT(QMC) results in Fig.~\ref{wnfig}, e.g.,
$d\!=\!w(f^2)=0.108\pm 0.008$ at $T\!=\!0.027\,$eV (316 K).

There are some differences between the three calculations,
however, which might account for differing $w(f^n)$ predictions:
(i) Our calculated Coulomb interaction for $\alpha$-Ce,
$U_f\!=\!5.7\,$eV, is slightly smaller than the $U_f\!=\!6\,$eV
employed in Refs.~\onlinecite{Zoelfl01} and \onlinecite{Liu92}.
(ii) Liu {\it et al.} employ an impurity Anderson model whereas
both we and Z\"olfl {\it et al.} extract a periodic Anderson type
of model from the LDA, including $f$-$f$ hybridization.  While we
also deal with an Anderson impurity model in the course of our DMFT
solution, this impurity model is only an auxiliary construction
with a complicated and strongly renormalized (non-constant)
hybridization.  (iii) Finally, in contrast to the DMFT(QMC),
both DMFT(NCA)\cite{Zoelfl01} and the $1/N$ approach\cite{1N} of
Ref.~\onlinecite{Liu92} are based on perturbation expansions in the
hybridization strength, a quantity which gets larger with reduced
volume.  Thus, while these two approximations are controlled by
the smallness of the hybridization strength and also by $1/N$
(we have $N\!=\!14$ $f$ orbitals), there are nonetheless larger
corrections when the hybridization is increased, i.e., when going
to the more itinerant $\alpha$-Ce.  Note in this context that the
ratio of Coulomb interaction to an effective bandwidth determined
by the total $f$-$f$ and $f$-valence hybridization changes from
3.8 to 2.5 across the $\gamma$-$\alpha$ transition.\cite{JCAMD}

It is possible to quantify the degree of $f$-electron
correlation by noting certain limiting values of $d$.
A natural minimum is provided by the strongly correlated
ground state of Eq.~(\ref{Ham}) in the atomic limit, where $d$
is a piecewise linear function of $n_f$, with $d\!=\!d_{\rm
min}=\max(0,n_f\!-\!1)$ for $n_f\!\leq\!2$.  Similarly, $d_{\rm
max}=(13/28)n_f^2$ from Eq.~(\ref{double}) in the uncorrelated
limit $\langle \hat{n}_1 \hat{n}_2\rangle=\langle \hat{n}_1
\rangle \langle \hat{n}_2\rangle$, which is approached for volume
$V\!\rightarrow\! 0$ leading to a vanishing ratio of Coulomb
interaction to bandwidth.  Fig.~\ref{dratfig} shows a plot of the
ratio $(d_{\rm max}\!-\!d)/(d_{\rm max}\!-\!d_{\rm min})$ for the
present Ce calculations at $T\!=\!0.054\,$eV, which reflects strong
and weak correlation limits at $1$ and $0$, respectively.  Note the
polarized to paramagnetic HF transition at $V\!\sim\!20\,$\AA$^3$
for decreasing volume, and the fact that the paramagnetic HF result
is completely uncorrelated ($d \approx d_{\rm max}$) as expected.
The fact that the $d$ ratio in this case is not precisely zero
is due to a small amount of orbital polarization arising from
the fact that $4f$ bands of different symmetry overlap the Fermi
level to slightly different extent, whereas $d_{\rm max}$ was
defined for all spin-orbital occupations to be $n_f/14$.

\begin{figure}[tb]
 \includegraphics[width=3.2in]{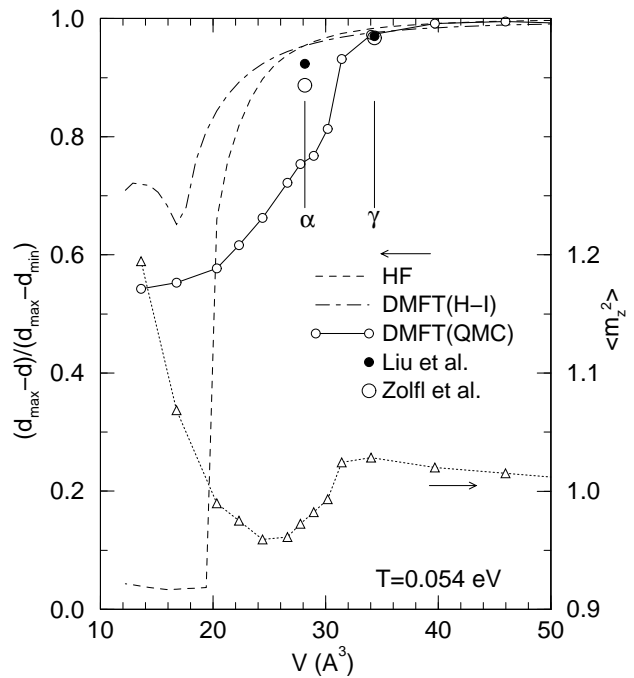}
\caption{Double occupation  ratio $(d_{\rm max}\!-\!d)/(d_{\rm
max}\!-\!d_{\rm min})$ and local magnetic moment $\langle
m_Z^2\rangle$ (triangles) as a function of volume at
$T\!=\!0.054\,$eV.  In the former case, we compare the
LDA+DMFT(QMC) results with our HF and LDA+DMFT(H-I) results
as well as with the LDA+ DMFT(NCA) by Z\"olfl {\it et al.}
\cite{Zoelfl01} and the Anderson model calculations by Liu et
al.\cite{Liu92} The double occupancy increases when going from
$\gamma$- to $\alpha$-Ce (experimental volumes as indicated),
i.e., the electrons become more itinerant or less correlated.
This effect is most pronounced in our LDA+DMFT(QMC) results; however,
the $d$ ratio is still far from the uncorrelated value $d=d_{\rm
min}$, i.e., $\alpha$-Ce is still strongly correlated.
\label{dratfig}}
\end{figure}

The combination of increasing $d$ and decreasing $n_f$
causes a sharp decrease in correlation (delocalization)
of the DMFT(QMC) result for decreasing volume through the
observed $\gamma$-$\alpha$ transition (marked), in agreement
with tenets of the Mott-transition model.\cite{JOHANSSON,HLMO}
The value of the DMFT(QMC) $d$ ratio is $0.76\pm0.08$ at the
$\alpha$ volume, combining all of the uncertainties in both $d$
and $n_f$.  While this value is certainly less correlated than the
DMFT(NCA)\cite{Zoelfl01} (large open circles) and impurity Anderson
model\cite{Liu92} (filled circles) predictions at 0.89 and 0.92,
respectively, it is far from the kind of uncorrelated behavior seen
in the paramagnetic HF of Fig.~\ref{dratfig} or, presumably also,
in the LDA.  Even at the smallest volumes considered, the DMFT(QMC)
$d$ ratio still suggests the presence of significant correlation,
which is entirely consistent with the remnant Hubbard side bands
in this range as discussed in the previous section.  Most notable
in the  DMFT(H-I) curve is a glitch at about $V\!=\!17\,$\AA$^3$
which is a consequence of the behavior in $n_f$ (not shown):
Within DMFT(H-I), $n_f$ is pinned at $1$ for decreasing volume
until $V\!=\!17\,$\AA$^3$, at which point it increases and the
system becomes mixed valent.

Turning to the local magnetic moment, our approximations
[neglect of spin orbit, intra-atomic exchange, and the $4f$
crystal field splitting in Eq.~(\ref{impurity})] have
more serious implications for this quantity than others,
and so we can provide only an estimate.  Consistent with
these approximations we take 
\begin{equation}
\langle \hat{n}_{ifm\sigma}\hat{n}_{ifm^\prime\sigma^\prime}
\rangle = \left\{ \begin{array}{ll} n_f/14 & \mbox{if $m\sigma
=m^\prime\sigma^\prime$} \\ d/91 & \mbox{if $m\sigma\neq
m^\prime\sigma^\prime$ ,} \end{array} \right.  
\label{n1n2eqn}
\end{equation}
such that the local magnetic moment becomes
\begin{equation} 
\langle m_z^2\rangle \equiv \sum_m \langle
(\hat{n}_{i f m \uparrow}\!-\!\hat{n}_{i f m\downarrow})^2\rangle = n_f -(2/13)\,d,
\label{lmomeqn}
\end{equation}
indicating whether a local spin moment exists.  Note that this
quantity does not contain information about long-range magnetic
order, aside from the fact that a finite moment would be required
for such order.  Also note that $\langle m_z^2\rangle$ is unlikely
to vanish.  Even if one just statistically distributes electrons
with arbitrary $i$, $m$, and $\sigma$, some sites will have
electrons with the same spin and thus $\langle m_z^2\rangle$
will be finite, but it will be smaller than its maximal value
obtained in the localized regime where $d$ is minimal.

The spin, orbital, and total angular momentum
expectations can be expressed as $\langle \hat{S}_{if}^2
\rangle\!=\!(3/4)\langle m_z^2\rangle$, $\langle \hat{L}_{if}^2
\rangle\!=\!12\langle m_z^2\rangle$, and $\langle \hat{J}_{if}^2
\rangle\!=\!(51/4)\langle m_z^2\rangle$ due to the degeneracies
in Eq.~(\ref{n1n2eqn}).  Note that in the atomic limit
($n_f\!\sim\! 1$, $d\!\sim\! 0$) these expressions correctly
give $S_{if}\!=\!1/2$ and $L_{if}\!=\!3$, although $\langle
\hat{J}_{if}^2 \rangle$ averages over the two spin-orbit
multiplets.  Our DMFT(QMC) result for $\langle m_z^2\rangle$ at
$T\!=\!0.054\,$eV is also provided in Fig.~\ref{dratfig} (bottom
dotted curve with open triangles), where this quantity is seen to
drop by 5\% from the $\gamma$ to the $\alpha$ volume.  This may be
compared to 11\% and 12\% drops for the DMFT(NCA)\cite{Zoelfl01}
and impurity Anderson model\cite{Liu92} calculations, respectively,
based on their values of $n_f$ and $d\!=\!w(f^2)$.  High-energy
neutron scattering experiments observe single-ion magnetic
response from $0.8\pm 0.1$ $4f$ electrons in the $\alpha$ phase,
suggesting also that much of the local moment persists into
that phase.\cite{Murani93} Such high-energy or ``fast'' probes
can detect a local moment even if it appears screened out in
``slower'' measurements like magnetic susceptibility.  Note that
the, at first view unexpected, increase in the DMFT(QMC) $\langle
m_z^2\rangle$ for the smallest volumes in Fig.~\ref{dratfig}
only reflects this same behavior in $n_f$ (Fig.~\ref{nffig}).

The persistence of a still robust (albeit slightly reduced)
local $4f$ moment into the $\alpha$ phase as suggested here
supports the Kondo Volume Collapse scenario,\cite{KVC}
in that the observed temperature-independent Pauli-like
paramagnetism of the $\alpha$ phase can then arise when
the valence electrons screen out these local moments.
Orbitally polarized\cite{Eriksson90,Soderlind02,Svane97}  and
self-interaction corrected\cite{Svane97,Svane94,Szotek94} LDA
results suggest that the moment actually collapses to nothing
in the $\alpha$ phase of Ce and its analog in Pr. However,
these calculations really measure spin and orbital polarization
analogous to $\langle m_z\rangle$, and therefore describe a loss
of magnetic order in the $\alpha$-like phases without providing
information about the local moment itself. Indeed, there can be a
local moment $\langle m_z^2 \rangle$
even in the fully uncorrelated limit, as noted
earlier, since $\langle m_z^2 \rangle = n_f\!-\!(2/13)d_{\rm
max} = n_f(1\!-\!n_f/14)$ can be significant away from empty
or filled bands.
Note that one may have temperature-independent paramagnetism in
the presence of local moments both if there is correlated Kondo
screening of these moments, as noted above, as well as by Pauli's
original one-electron process in which only electrons
in states near the Fermi level are free to respond to the field.
The latter must dominate as one approaches the uncorrelated
$V\!=\!0$ limit.

\section{Summary and Discussion}

We have calculated thermodynamic, spectral, and other properties
of Ce metal over a wide range of volumes and temperatures using
the merger of the local density approximation and dynamical mean
field theory (LDA+DMFT).  The DMFT self energy was generated by
rigorous quantum Monte Carlo (QMC) techniques, including a new,
faster implementation that has facilitated lower-temperature
results and is described in detail.   Our LDA+DMFT results provide
a comprehensive picture of correlation effects in compressed Ce,
and their fundamental role in the first-order $\alpha$-$\gamma$
transition. First results of this effort have been published in
Ref.~\onlinecite{Held01b}.

At large volume, we find a Hubbard split $4f$ spectrum, the
associated local magnetic moment, and an entropy reflecting the
degeneracy in the moment direction. This phase is well described
by the Hubbard-I approximation and its energy but not its entropy
also agrees with the polarized Hartree-Fock solution.  As volume
is reduced, a quasiparticle or Abrikosov-Suhl resonance begins to
develop at the Fermi level in the vicinity of the $\alpha$-$\gamma$
transition, and the entropy starts to drop.  At the same time, the
 $4f$ double occupation grows whereas the number of $4f$
electrons remains close to one.  The temperature dependence of
the quasiparticle peak is consistent with a significantly larger
Kondo temperature in the $\alpha$ phase than in the $\gamma$
phase, and the parameter-free  LDA+DMFT spectra are in good
agreement with experiment for both $\alpha$- and $\gamma$-Ce.
In the range where the quasiparticle peak grows dramatically,
the correlation energy as a function of volume is seen to have a
negative curvature.  This leads to a growing shallowness in the
total energy as temperature is reduced and is consistent with the
first-order $\alpha$-$\gamma$ transition within our error bars.
Our results suggest that the temperature dependence of the
transition pressure is  primarily due to the entropy.  Finally,
if the volume is reduced below that of the ambient $\alpha$ phase,
the quasiparticle peak grows at the expense of the Hubbard side
bands, yet these Hubbard side bands persist even at the smallest
volumes  considered.

The Mott transition\cite{Mott} (MT) and Kondo volume
collapse\cite{KVC} (KVC) scenarios are based on the one-band
Hubbard and the periodic (or more approximate impurity)
Anderson models as paradigms.  The classification of our
results in terms of these two standard theories requires
distinguishing between the more general interpretation of the
MT in the many body community,\cite{Gebhard97} e.g., applied
to such materials as V$_2$O$_3$,\cite{Held01a} and the ideas
of Johansson\cite{JOHANSSON} and members of the local-density
functional community as applied to the $f$-electron metals.
\cite{Eriksson90,Soderlind02,Svane97,Svane94,Szotek94,Sandalov95,Shick01}
In the former case, correlated solutions of {\it both} model
Hamiltonians show similar features at finite temperature such
as persistence into the more weakly correlated regime of the
local moment and residual Hubbard splitting,\cite{Held00ab}
just as seen here for $\alpha$-Ce.  The similarity between the
two models can be understood from the following consideration:
The conduction-electrons in the periodic Anderson model are
non-interacting.  Thus, they only enter quadratically in the
effective action and can be integrated out by a simple Gauss
integration.  This results in an effective one-band model for
the $f$ electrons of the periodic Anderson model which can behave\cite{Held00ab}
very much like the Hubbard model not only at finite temperature,
but, depending on the choice of $f$-$d$ hybridization, also at
$T\!=\!0$. 

One might try to distinguish between the two scenarios
by whether the transition is caused by changes of the $f$-$f$
(MT) or $f$-valence (KVC) hybridization.
But, since realistic calculations like the present include both,
this distinction is rather problematic.
Another difference can be addressed unambiguously,
i.e., whether
the low-temperature $\gamma$ phase 
has an  Abrikosov-Suhl resonance (KVC) or not (MT).
We observe the former, in agreement with experiment.\cite{Liu92} 
The energy scale of this $\gamma$-Ce Abrikosov Suhl resonance
is very small such that we obtain a thermally smeared out
structure instead of a sharp resonance. The smallness of the
energy scale also implies that the effect on 
the total energy is  very minor.
 Because of this, the low-temperature energy
(but not the entropy) of $\gamma$-Ce may also be adequately
described by static mean field techniques like our HF calculation
as well as a number of local-density functional modifications:
orbitally polarized LDA,\cite{Eriksson90,Soderlind02,Svane97}
self-interaction corrected LDA,\cite{Svane97,Svane94,Szotek94}
and LDA+U.\cite{Sandalov95,Shick01} These approximations have
a frequency-independent (static) self-energy and provide a
splitting of the $4f$-band into two bands by an (artificial)
symmetry breaking.  While our HF calculations as well as LDA+U
work\cite{Sandalov95} for Ce give a transition at too small
volume, one may drive the onset of the symmetry breaking closer
to the volume of the $\alpha$-$\gamma$ transition by reducing the
the $4f$ Coulomb interaction $U_f$.  Such a reduced interaction
strength is naturally achieved within the orbitally polarized LDA
calculations which omit $U_f$ and employ the weaker intra-atomic
exchange interaction to achieve the symmetry breaking.

A major point of debate between the KVC scenario and Johansson's
interpretation of the MT picture is whether the $\alpha$ phase
of Ce is strongly correlated (KVC) or not (MT).  Our results
suggest that $\alpha$-Ce is strongly correlated with a three
peak structure consisting of the two Hubbard peaks and central
quasiparticle bands as in the KVC picture.\cite{KVC} In contrast,
the MT model as advocated by Johansson\cite{JOHANSSON} and others
\cite{Eriksson90,Soderlind02,Svane97,Svane94,Szotek94,Sandalov95}
predicts a single peak associated with uncorrelated, band-like $f$
electrons.  We do see a rapid increase in double occupation $d$
across the transition, which is consistent with the delocalization
ideas of this MT scenario.  However, the actual value of $d$ in
the $\alpha$ phase is far from uncorrelated, although it indicates
considerably less correlation than in the KVC picture.\cite{Liu92}
It appears that this perspective of the MT is motivated by
the LDA results, and that if one were to fully take into account
electronic correlations, one would also observe a correlated
three-peak solution as in the Hubbard model.\cite{Held00ab} This
correlated solution would also have preformed local magnetic
moments, which would be screened at small energies on the scale
of the width of the  Abrikosov-Suhl resonance, as in the KVC
picture, whereas the uncorrelated MT solution does not develop
such local moments.

Since we find a strongly correlated $\alpha$ phase, the
question remains why the structural and volume dependence of
the total energy in the $\alpha$-Ce regime is so extremely
well described by normal paramagnetic LDA and its gradient
corrected improvements.\cite{Brooks84,Soderlind98}
This point is one of the strongest arguments
advanced by Johansson\cite{JOHANSSON} and others
\cite{Brooks84,Soderlind98,Eriksson90,Soderlind02,Svane97,Svane94,Szotek94,Sandalov95}
that $\alpha$-Ce-like phases should be weakly correlated.
A logical explanation would be that LDA may get the interactions
between the quasiparticles correct but not their formation
energy.  The interactions are perhaps governed by the significant
weight in the central Fermi-level peak which resembles the
uncorrelated LDA solution, while the formation energy may involve
the residual Hubbard sidebands in some way.  Thus the still very
significant correlation may provide only a constant contribution
to the total energy in the $\alpha$-Ce regime, so that the
volume and structural dependence is still well represented.
This would be consistent with the energy shift between $\alpha$
and $\gamma$ phases employed by Johansson et al.\cite{JOHANSSON2}
in their LDA-based modeling of the transition.

\section*{Acknowledgments}

Work by AKM was performed under the auspices of the
U. S. Department of Energy by the University of California,
Lawrence Livermore National Laboratory under contract
No. W-7405-Eng-48.  KH acknowledges support by the Alexander
von Humboldt foundation, and RTS from NSF-DMR-9985978.  We are
grateful for the QMC code of Ref.~\onlinecite{DMFT2} (App.~D)
which was modified for use in part of the present work.  We thank
A.~Sandvik for making available his maximum entropy code, I.~A.~Nekrasov 
for providing the digitized experimental spectra, and
G.~Esirgen, J. W.~Allen, and  O.~Gunnarsson for useful discussions.


\begin{thebibliography}{99}
\bibitem{dtauU} 
We cite $\Delta\tau U/2$ values for $U_f\!=\!5.81\,$eV at
$V\!=\!30\,$\AA$^3$, while $U_f$ is 4.92 and $6.27\,$eV at $V=15$
and $45\,$\AA$^3$, respectively.
\bibitem{NoteSpectra}
The $f$-electron Coulomb interactions have been obtained by a
constrained LDA calculation and the experimental resolution has
been taken from Ref.~\onlinecite{Zoelfl01}.
\bibitem{nfmeas}
These vaues are averages of $n_f$ from Eq.~(\ref{nf}) and $n^{\rm
QMC}_f=14[1\!+\!G(\tau\!=\!0^+)]$ from the QMC.
\bibitem{HLMO}
The intuitive idea is captured by the Heitler-London ($d\!=\!0$),
$[\phi_a(1)\phi_b(2)+\phi_b(1)\phi_a(2)]/\sqrt{2}$,
versus molecular orbital ($d\!=\!1/4$),
$[\phi_a(1)\!+\!\phi_b(1)][\phi_a(2)\!+\!\phi_b(2)]/2$, spatial
wavefunctions for the hydrogen molecule (sites $a$ and $b$).
These represent completely correlated (localized) and completely
uncorrelated (itinerant) ground states, respectively.  The per site
$d$ values agree with the general expressions for $d_{\rm min}$
and $d_{\rm max}$ in the text, noting that the coefficient of
$n_f^2$ in the latter is $(2z\!-\!1)/(4z)$ for $z$ ``$f$'' orbitals
per site, and $z\!=\!n_f\!=1$ for the hydrogen example.
\end{thebibliography}
\end{document}